\documentclass[12pt]{article}
\usepackage{epsfig,amsmath,amssymb}
\usepackage{graphicx,psfrag} 
\usepackage{mathtools}
\hypersetup{linktocpage}
\numberwithin{equation}{section}

\newcommand{\be}{\begin{equation}}
\newcommand{\ee}{\end{equation}}
\newcommand{\bea}{\begin{eqnarray}}
\newcommand{\eea}{\end{eqnarray}}
\newcommand{\bA}{\begin{array}}
\newcommand{\eA}{\end{array}}
\newcommand{\bc}{\begin{center}}
\newcommand{\ec}{\end{center}}

\newcommand{\ra}{\rightarrow}
\newcommand{\del}{\partial}

\newcommand{\ie}{{\it i.e.}}
\newcommand{\eg}{{\it e.g.}}

\begin{document}

\begin{titlepage}
	
	\bc

	\hfill 
	\\         [22mm]

	{\Huge $AdS_2$ dilaton gravity from\\ 
		[2mm]
		reductions of some nonrelativistic theories}
	\vspace{16mm}
	
	{\large Kedar S. Kolekar,\ \ K. Narayan}\\
	\vspace{3mm}
	{\small \it Chennai Mathematical Institute, \\ }
	{\small \it SIPCOT IT Park, Siruseri 603103, India.\\ }
	
	\ec
	\medskip
	\vspace{30mm}
	
\begin{abstract}
We study dilaton-gravity theories in 2-dimensions obtained by
dimensional reduction of higher dimensional nonrelativistic
theories. Focussing on certain families of extremal charged
hyperscaling violating Lifshitz black branes in
Einstein-Maxwell-scalar theories with an extra gauge field in
4-dimensions, we obtain $AdS_2$ backgrounds in the near horizon
throats. We argue that these backgrounds can be obtained in equivalent
theories of 2-dim dilaton-gravity with an extra scalar, descending
from the higher dimensional scalar, and an interaction potential with
the dilaton. A simple subcase here is the relativistic black brane in
Einstein-Maxwell theory. We then study linearized fluctuations of the
metric, dilaton and the extra scalar about these $AdS_2$
backgrounds. The coefficient of the leading Schwarzian derivative term
is proportional to the entropy of the (compactified) extremal black
branes.
\end{abstract}

\end{titlepage}
        
\newpage
{\footnotesize
\tableofcontents
}

\vspace{-2mm}
        
\section{Introduction}

Gravity in two dimensions, trivial as such, is rendered dynamical in
the presence of a dilaton scalar and additional matter. Such
dilaton-gravity theories arise generically under dimensional reduction
from higher dimensional theories of gravity coupled to matter. There
is interesting interplay with $AdS_2$ holography, which arises in the
context of extremal black holes and branes: the near horizon regions
typically acquire an $AdS_2\times X$ geometry, and a 2-dimensional
description arises after compactifying the transverse space $X$.
Almheiri and Polchinski \cite{Almheiri:2014cka} considered toy models
of 2-dim dilaton-gravity of this sort, with backgrounds involving
$AdS_2$ with a varying dilaton. Analyzing the backreaction of a
minimally coupled scalar perturbation on the $AdS_2$ background
reveals nontrivial scaling of boundary 4-point correlation functions
thereby indicating the breaking of $AdS_2$ isometries in the deep
IR. This breaking amounts to breaking of local reparametrizations of
the boundary time coordinate (modulo global $SL(2)$ symmetries), which
would have been preserved in the presence of exact conformal symmetry.
In \cite{Maldacena:2016upp}, as well as \cite{Jensen:2016pah,
Engelsoy:2016xyb,Almheiri:2016fws}, it was argued that the leading effects
describing such nearly $AdS_2$ theories are captured universally by a
Schwarzian derivative action governing boundary time
reparametrizations modulo $SL(2)$, which arises from keeping the
leading nonconstant dilaton behaviour. This picture dovetails with
the absence of finite energy excitations in $AdS_2$ discussed
previously in \cite{Maldacena:1998uz,Sen:2008vm}.
Parallel exciting developments involve various recent investigations
of the SYK model \cite{Sachdev:1992fk}, \cite{Kitaev-talks:2015},
\cite{Kitaev:2017awl}, a quantum mechanical model of interacting
fermions. This exhibits approximate conformal symmetry at low
energies: the leading departures from conformality are governed by
a Schwarzian derivative action for time reparametrizations
modulo $SL(2)$, as above. A recent review is \cite{Sarosi:2017ykf}.

$AdS_2$ throats arise quite generally in the near horizon regions of
extremal black holes and black branes, where other fields acquire near
constant ``attractor'' values. This attractor mechanism, first
discussed in \cite{Ferrara:1995ih} for BPS black holes in $N=2$
theories, arises from extremality rather than supersymmetry, as
studied in \cite{Goldstein:2005hq}, \cite{Tripathy:2005qp}. In the
last several years, this has been ubiquitous in the context of
nonrelativistic generalizations of holography: a nice review is
\cite{Hartnoll:2016apf}. A large family of such theories is obtained
by considering Einstein-Maxwell-scalar theories with a negative
cosmological constant and potential: the $U(1)$ gauge field and scalar
serve to support the nonrelativistic background, typically of the form
of a Lifshitz, or hyperscaling violating (conformally Lifshitz)
theory.  The duals to the bulk uncharged black branes in these hvLif
theories capture many features of finite density condensed matter-like
systems. Towards studying extremal black branes, we note that charge
can be added to these theories by adding an additional $U(1)$ gauge
field, as discussed in \eg\ \cite{Tarrio:2011de},
\cite{Alishahiha:2012qu}, \cite{Bueno:2012sd}. Now at extremality, the
infrared region approaches an $AdS_2\times X$ throat, with $X$
typically of the form of an extended transverse plane $R^d$. The
discussion above of $AdS_2$ holography now applies upon compactifying
$X$ taken as \eg\ a torus $T^d$.\ This was in fact the broad context
for \cite{Almheiri:2014cka}:\ other recent discussions of reduction
from higher dimensional theories appear in
\eg\ \cite{Cvetic:2016eiv,Castro:2008ms,Castro:2014ima,Das:2017pif,Taylor:2017dly,Gaikwad:2018dfc,Nayak:2018qej};\ see also \cite{Cadoni:2017dma}.

Towards studying such $AdS_2$ theories arising in this nonrelativistic
context, we study effective gravity theories of the above form, with
two $U(1)$ gauge fields and a scalar field $\Psi$ with a negative
cosmological constant and potential. We focus for concreteness on the
charged hyperscaling violating Lifshitz black branes in $4$-dimensions
described in \cite{Alishahiha:2012qu}. In the extremal limit, the near
horizon geometry of these charged hyperscaling violating Lifshitz
black branes becomes $AdS_2\times \mathbb{R}^2$. These charged
hyperscaling violating Lifshitz attractors arise for certain regimes
of the Lifshitz $z$ and hyperscaling violating $\theta$ exponents
allowed by the energy conditions, with the additional requirement that
the theory exhibits hvLif boundary conditions in the ultraviolet:
these are perhaps best regarded as intermediate infrared phases
themselves in some bigger phase diagram. Then compactifying the two
spatial directions as a torus $T^2$, we dimensionally reduce this
charged hvLif extremal black brane to obtain a $2$-dimensional
dilaton-gravity-matter theory. This theory is equivalent to gravity
with a dilaton $\Phi$ and an additional scalar $\Psi$ that descends
from the hvLif scalar in the higher dimensional theory, along with an
interaction potential $U(\Phi,\Psi)$. The interaction potential raises
the question of whether the extra scalar destabilizes the $AdS_2$
regime, possibly in some region of parameter space. Towards
understanding this, we study small fluctuations about the extremal
$AdS_2$ background in these theories and argue that these are in fact
stable, the stability stemming from the restrictions imposed on
$z,\theta$ stated above from energy conditions and asymptotic boundary
conditions. Studying the action for small fluctuations up to quadratic
order, it can be seen that the leading corrections to $AdS_2$ arise at
linear order in $\delta\Phi$ leading again to a Schwarzian derivative
action from the Gibbons-Hawking term, although there are subleading
coupled quadratic corrections (sec.~\ref{sec:chvLif}). The coefficient
of the Schwarzian is proportional to the entropy of the compactified
extremal black branes, which being the number of microstates of the
background is akin to a central charge of the effective theory. In
sec.\ref{sec:relElecbrane}, we first describe in detail the simpler
case of the relativistic black brane, which has $z=1, \theta=0$,
arising in Einstein-Maxwell theory, the extra scalar being absent: 
at leading order this shows how the Jackiw-Teitelboim theory
\cite{Jackiw:1984je,Teitelboim:1983ux} arises, with subleading terms
at quadratic order. We finally study in sec.~5 a null reduction of the
charged relativistic black brane: this results in charged hvLif black
brane backgrounds with specific exponents, but with an extra scalar
background profile (for the uncharged case, these coincide with
\cite{Narayan:2012hk}). Sec.~6 contains a brief Discussion and an
Appendix contains some technical details.

\section{Einstein-Maxwell theory in $4$-dimensions}

Einstein-Maxwell theory with a negative cosmological constant is a
useful playground for various interesting physics: see
\eg\ \cite{Hartnoll:2016apf} for a review. We focus on 4-dimensions
for simplicity: as a consistent truncation of M-theory on appropriate
7-manifolds, the bulk gauge field can be taken as the dual to the
$U(1)_R$ current. The action is
\begin{equation}\label{cbb4daction}
	S=\int d^4x\sqrt{-g^{(4)}}\left[\frac{1}{16\pi G_4}\Big(\mathcal{R}^{(4)}-2\Lambda\Big)-\frac{1}{4}F_{MN}F^{MN}\right]\ ,
\end{equation}
where $\Lambda=-3$ is the cosmological constant in $4$-dimensions. The field equations are
\begin{equation}\label{cbb4deoms}
	\mathcal{R}^{(4)}_{MN}-\Lambda g_{MN}-8\pi G_4\Big(F_{MP}F_{N}^{\ P}-\frac{g_{MN}}{4}F^2\Big)=0\ , \qquad \partial_{M}(\sqrt{-g}F^{MN})=0\ .
\end{equation}
These equations have both electrically and magnetically charged black
branes as solutions.

\vspace{1mm}

\noindent \underline{\emph{Magnetic branes:}}\ \ These are slightly
simpler and we discuss them first, mostly reviewing discussions already
in the literature. The metric and field strength \cite{DHoker:2009mmn} are
\bea\label{cbb4dsolnm}
ds^2=-r^2 f(r)dt^2 + \frac{dr^2}{r^2f(r)} + r^2(dx^2+dy^2)\ ,  &&
f(r)=1-\left(\frac{r_0}{r}\right)^3 +
\frac{Q_m^2}{r^4}\Big(1-\frac{r}{r_0}\Big)\ , \nonumber\\ [2pt]
F_{xy}=Q_m\ , &&
\eea
where $Q_m$ is related to the magnetic charge of the black brane, $r_0$ is
the location of the horizon and $r\rightarrow\infty$ is the
boundary. In the extremal limit, the Hawking temperature vanishes, fixing
the horizon location in relation to the charge,
\begin{equation}\label{cbb4dextremalitym}
  T=\frac{3r_0}{4\pi}\Big(1-\frac{Q_m^2}{3r_0^4}\Big)=0 \quad
  \implies\ Q_m^2=3r_0^4\ .
\end{equation}
The near horizon geometry of the magnetic black brane becomes
$AdS_2\times \mathbb{R}^2$,
\begin{equation}\label{cbb4dexads2m}
  ds^2=-r_0^2f(r)dt^2 + \frac{dr^2}{r_0^2f(r)} + r_0^2(dx^2+dy^2)\ ,
  \qquad f(r)|_{r\rightarrow r_0}\simeq \frac{6}{r_0^2}(r-r_0)^2\ .
\end{equation}
We compactify the two spatial dimensions $x^i$ as $T^2$ and
dimensionally reducing with an ansatz for the metric
\begin{equation}\label{compactmetric}
	ds^2=g^{(2)}_{\mu\nu}dx^{\mu}dx^{\nu} +\Phi^2(dx^2+dy^2)\ ,
\end{equation}
with $g^{(2)}_{\mu\nu}$ and $\Phi$ being independent of the compact
coordinates $x,y\in T^2$.\
The action \eqref{cbb4daction} for the magnetic black brane solution
then reduces to
\begin{equation}\label{cbb2dactionm}\begin{split}
S=\frac{1}{16\pi G_2}\int d^2x\sqrt{-g^{(2)}}\, \Big[\Phi^2\mathcal{R}^{(2)}-2\Lambda\Phi^2-\frac{Q_m^2}{2\Phi^2}+2\partial_\mu\Phi\partial^\mu\Phi\Big] \ ,
\end{split}\end{equation}
where $G_2=G_4/V_2$ is the dimensionless Newton constant in 2-dimensions.\
A Weyl transformation $g_{\mu\nu}=\Phi g^{(2)}_{\mu\nu}$ absorbs
the kinetic term for the dilaton $\Phi$ in the Ricci scalar giving
\begin{equation}\label{cbb2dactionmWeyl}
S=\frac{1}{16\pi G_2}\int d^2x\sqrt{-g}\,\Big(\Phi^2\mathcal{R}-2\Lambda\Phi-\frac{Q_m^2}{2\Phi^3}\Big)\ \equiv\ \frac{1}{16\pi G_2}\int d^2x\sqrt{-g}\left(\Phi^2\mathcal{R}-U(\Phi)\right) .
\end{equation}
The equations of motion from this action are
\begin{eqnarray}\label{cbb2deeomsm}
U(\Phi) = 2\Lambda\Phi+\frac{Q_m^2}{2\Phi^3}\ ;&& \qquad\qquad 
\mathcal{R}- {\del U\over\del\Phi^2} = 0\ ,\nonumber\\[3pt]
&& g_{\mu\nu}\nabla^2\Phi^2-\nabla_{\mu}\nabla_{\nu}\Phi^2
+ \frac{g_{\mu\nu}}{2}\, U(\Phi) = 0\ .
\end{eqnarray}	
This $2$-dimensional dilaton-gravity theory admits $AdS_2$ as a
solution with a constant dilaton. This constant dilaton, $AdS_2$
solution is just the near horizon $AdS_2$ geometry of the extremal
magnetic black brane in $4$-dimensions (which asymptotically, as
$r\ra\infty$, is $AdS_4$).

The purpose of this section was to simply illustrate that the original
theory with the gauge field is equivalent to a dilaton-gravity theory
with an appropriate dilaton potential: this will be a recurrent theme.
A simple toy model capturing many features of 2-dim dilaton gravity
is the Jackiw-Teitelboim theory \cite{Jackiw:1984je,Teitelboim:1983ux}.
In the discussion above, we have not been careful with length-scales:
in the next section for the relativistic electric brane, we will
reinstate various scales.

\subsection{Relativistic electric black brane,\ reduction to $2$-dimensions}
\label{sec:relElecbrane}

The electric black brane solution to (\ref{cbb4daction}), \eqref{cbb4deoms}, is
\bea\label{cbb4dsolne}
ds^2=-\frac{r^2f(r)}{R^2}dt^2+\frac{R^2}{r^2f(r)}dr^2+\frac{r^2}{R^2}(dx^2+dy^2)\ , &&
f(r)=1-\left(\frac{r_0}{r}\right)^3+\frac{Q_e^2}{r^4}\Big(1-\frac{r}{r_0}\Big)\ , \nonumber\\ [2mm]
A_t=\frac{Q_e}{2\sqrt{\pi G_4}\,Rr_0}\Big(1-\frac{r_0}{r}\Big)\ , &&
F_{rt}=\frac{Q_e}{2\sqrt{\pi G_4}\,R}\,\frac{1}{r^2}\ .
\eea
The gauge field $A_t$ vanishes at the horizon.
The charge parameter $Q_e$ is related to the chemical potential $\mu$ and
the charge density $\sigma$ of the black brane as
\begin{equation}\label{cbb4dQemurho}
  {Q_e\over 2\sqrt{\pi G_4}\,Rr_0} = \mu\ , \qquad
  \sigma=\mu\frac{r_0}{R^2}=\frac{Q_e}{2\sqrt{\pi G_4}\,R^3}\ .
\end{equation}
Reinstating the dimensionless gauge coupling $e^2$ in $\mu$ and
$\sigma$ as $\mu\rightarrow \frac{\mu}{e}$ and $\sigma\rightarrow
\sigma e$ and using \eqref{cbb4dQemurho}, we recover the expressions
for the gauge field, field strength and the thermal factor in terms of
$r_0$, $\mu$, $\sigma$ as given in sec.~4.2.1 in
\cite{Hartnoll:2016apf}.\ Note that in (\ref{cbb4dsolne}) the charge
parameter $Q_e$ has dimensions of charge times length-squared, and the
gauge field $A_t$ has mass dimension one.\ 
In the extremal limit, the temperature vanishes giving
\begin{equation}\label{cbb4dextremalitye}
  T=\frac{3r_0}{4\pi R^2}\Big(1-\frac{Q_e^2}{3r_0^4}\Big)=0 \qquad
  \implies\qquad  Q_e^2=3r_0^4\ .
\end{equation}
The near horizon geometry of the electric black brane becomes
$AdS_2\times \mathbb{R}^2$, 
\begin{equation}\label{cbb4dexads2e}
  ds^2=-\frac{r_0^2}{R^2}f(r)dt^2+\frac{R^2}{r_0^2f(r)}dr^2+\frac{r_0^2}{R^2}(dx^2+dy^2)\ ,
  \qquad f(r)|_{r\rightarrow r_0}\simeq \frac{6}{r_0^2}(r-r_0)^2\ ,
\end{equation}
as in the magnetic case. The Bekenstein-Hawking entropy is the horizon
area in Planck units\
\be\label{entropyz=1}
S_{BH} = {r_0^2\over R^2}\, {V_2\over 4G_4}
= {Q_e/\sqrt{3}\over R^2}\, {V_2\over 4G_4}\ .
\ee
With $V_2=\int dx dy$ the area, this is finite entropy density for
noncompact branes.

It is worth noting that asymptotically, these branes (\ref{cbb4dsolne})
give rise to an $AdS_4$ geometry, with scale $R$. In the near horizon
region, we obtain an $AdS_2$ throat with scale ${R\over\sqrt{6}}$: this
is a well-defined $AdS_2$ throat in the regime\ ${r-r_0\over R} \gg 1$
and ${r-r_0\over r_0}\ll 1$\,.\
The $AdS_2$ region is well-separated from the boundary of the $AdS_4$
geometry at $r\sim r_C\gg r_0$\ if\ \ ${r-r_0\over r_c}\ll 1$.

Compactifying the two spatial dimensions $x^i$ as $T^2$ and dimensionally
reducing with the metric ansatz (\ref{compactmetric}) reduces the action
\eqref{cbb4daction} for the electric black brane solution to
\begin{equation}\begin{split}
	S=\int d^2x\sqrt{-g^{(2)}}\, \Big[\frac{1}{16\pi G_2}(\Phi^2\mathcal{R}^{(2)}-2\Lambda\Phi^2+2\partial_\mu\Phi\partial^\mu\Phi)-\frac{V_2\Phi^2}{4}F_{\mu\nu}F^{\mu\nu}\Big] \ ,
\end{split}\end{equation}
and we have suppressed a total derivative term which cancels with a
corresponding term arising from the dimensional reduction of the
Gibbons-Hawking boundary term (more on this later).\
Performing a Weyl transformation $g_{\mu\nu}=\Phi g^{(2)}_{\mu\nu}$ to absorb
the kinetic term for the dilaton $\Phi^2$ in the Ricci scalar, we get
\begin{equation}\label{cbb2dFaction}
	S=\int d^2x\sqrt{-g}\, \Big[\frac{1}{16\pi G_2}(\Phi^2\mathcal{R}-2\Lambda\Phi)-\frac{V_2\Phi^3}{4}F_{\mu\nu}F^{\mu\nu}\Big] \ .
\end{equation}
The Maxwell equations for the gauge field are
\begin{equation}\label{cbb2dFEWeyl}
	\partial_{\mu}(\sqrt{-g}\,\Phi^3 F^{\mu\nu})=0\ .
\end{equation}
The two components $b=t, r$ of \eqref{cbb2dFEWeyl}, \ie\
$\partial_t(\sqrt{-g}\Phi^3 F^{tr})=0=\partial_r(\sqrt{-g}\Phi^3 F^{tr})$,\
imply
\begin{equation}
	\sqrt{-g}\, \Phi^3 F^{tr}=const\ .
\end{equation}
Using the gauge field solution in \eqref{cbb4dsolne} to fix this constant
as $\frac{Q_e}{2\sqrt{\pi G_4}R^3}$, we get
\begin{equation}\label{cbb2dFsoln}
  F^{\mu\nu}=\frac{Q_e}{2\sqrt{\pi G_4}\,R^3}\frac{1}{\sqrt{-g}\,\Phi^3}\,\varepsilon^{\mu\nu}\ ,
\end{equation}
where $\varepsilon^{\mu\nu}$ is defined as
$\varepsilon^{tr}=1=-\varepsilon^{rt}$ and
$\varepsilon_{\mu\nu}=g_{\mu\rho}g_{\nu\sigma}\varepsilon^{\rho\sigma}$.
Substituting $F_{\mu\nu}F^{\mu\nu}=\frac{-Q_e^2}{2\pi G_4 R^6\Phi^6}$ and
$F_{\mu\rho}F_{\nu}^{\ \rho}=\frac{-Q_e^2}{4\pi G_4 R^6\Phi^6}g_{\mu\nu}$ in
eqns.\eqref{cbb2deoms}, we get
\begin{eqnarray}\label{cbb2deeoms}
  g_{\mu\nu}\nabla^2\Phi^2-\nabla_{\mu}\nabla_{\nu}\Phi^2+\frac{g_{\mu\nu}}{2}\Big(2\Lambda\Phi+\frac{2Q_e^2}{R^6\,\Phi^3}\Big)&=&0\ , \nonumber\\[3pt]
  \mathcal{R}-\frac{\Lambda}{\Phi}+\frac{3Q_e^2}{R^6\,\Phi^5}&=&0\ .
\end{eqnarray}
These field equations can be obtained by varying the following equivalent
action
\be\label{cbb2dequivaction}
S=\frac{1}{16\pi G_2}\int d^2 x\sqrt{-g}\,
\Big(\Phi^2\mathcal{R}-2\Lambda\Phi-\frac{2Q_e^2}{R^6\Phi^3}\Big)\ \equiv\
\frac{1}{16\pi G_2}\int d^2 x\sqrt{-g}\,
\Big(\Phi^2\mathcal{R}- U(\Phi)\Big) ,\
\end{equation}
This equivalent action is obtained by substituting the solution for
$F^{\mu\nu}$ (in terms of the dilaton $\Phi^2$) in the action
\eqref{cbb2dFaction} and changing the sign of the $F^2$ term which contains
a minus sign for electric branes alone, arising from $g_{tt}$\ (a similar
treatment appears also in \eg\ \cite{Almheiri:2016fws}).\ Note that this
is also consistent with and expected from electric-magnetic duality\
$Q_e\ra Q_m,\ Q_m\ra -Q_e$, which would suggest that the effective dilaton
potential for magnetic branes (\ref{cbb2dactionmWeyl}) is unchanged in
going to electric branes.\ Now for
instance the second equation in (\ref{cbb2deeoms}) becomes\
$R-{\del U\over \del\Phi^2} = 0$.\ The constant dilaton, $AdS_2$ solution
to the equations \eqref{cbb2deeoms}, consistent with the $T^2$
compactification of the near horizon geometry in (\ref{cbb4dexads2e}), is
\begin{equation}\label{cbb2dads2soln}
  ds^2=L^2\Big(-\frac{r_0^2}{L^4R^2}(r-r_0)^2dt^2+\frac{dr^2}{(r-r_0)^2}\Big)\ ,
  \qquad \Phi=\frac{r_0}{R}\ , \qquad L^2 = {Rr_0\over 6}\ ,
  \qquad Q_e^2=3r_0^4\ ,
\end{equation}
with $L$ the $AdS_2$ scale. Changing the
radial coordinate to $\rho=\dfrac{R^2}{6(r-r_0)}$\,, we write the metric
in conformal gauge
\begin{equation}\label{cbbconfgauge}\begin{split}
  ds^2&=e^{2\omega}(-dt^2+d\rho^2)=e^{2\omega}(-dx^+dx^-)\ , \qquad e^{2\omega}=\frac{L^2}{\rho^2}\ ,
\end{split}\end{equation}
where the lightcone coordinates are $x^{\pm}=t\pm\rho$. To see that
\eqref{cbb2dequivaction} admits the above $AdS_2$ solution, we compute
$\frac{\partial U}{\partial\Phi^2}$ for the above solution, which gives
\begin{equation}
  \frac{\partial U}{\partial\Phi^2}=-\frac{12}{Rr_0}=-\frac{2}{L^2}\qquad
\implies\qquad \mathcal{R}=\frac{\partial U}{\partial\Phi^2}=-\frac{2}{L^2}\ ,
\end{equation}
using \eqref{cbb2deeoms} for the Ricci scalar.  This constant dilaton,
$AdS_2$ solution (\ref{cbb2dads2soln}) is just the compactification of
the near horizon $AdS_2$ geometry of the 4-dim extremal electric black brane.

\subsubsection{Perturbations about the constant dilaton, $AdS_2$ background}

The 4-dimensional theory has a large spectrum of tensor, vector and
scalar perturbations, which upon reduction to 2-dimensions give a
corresponding spectrum: we will discuss this briefly later, in
sec.~\ref{sec:moregenpert}. In this section, we focus on perturbations
to only those fields that have nontrivial background profiles in the
effective 2-dimensional dilaton-gravity theory: thus we turn on
perturbations to the metric and the dilaton
\begin{equation}\label{cbb2dperturbations}
	\Phi=\Phi_b + \phi(x^+,x^-)\ , \qquad \omega=\omega_b+\Omega(x^+,x^-)\ ,
\end{equation}
where $\Phi_b$ and $\omega_b$ denote the background
\eqref{cbb2dads2soln}. We expand the action \eqref{cbb2dequivaction}
(in conformal gauge) about this background upto quadratic order to get
\begin{equation}
	S=\frac{1}{16\pi G_2}\int d^2x\, \Big(4\Phi^2\partial_+\partial_-\omega-\frac{e^{2\omega}}{2}U(\Phi)\Big)\  \equiv\ S_0+S_1+S_2\ ,
\end{equation}
where
\begin{equation}\label{S0z=1}
	S_0=\frac{1}{16\pi G_2}\int d^2x\, \Big(4\Phi_b^2\partial_+\partial_-\omega_b-\frac{e^{2\omega_b}}{2}U(\Phi_b)\Big)
\end{equation}
is the background action and $S_1$ is linear in perturbations and
vanishes by equations of motion. $S_2$ is quadratic in perturbations
given by
\begin{equation}\label{cbb2dS2}
		S_2=\frac{1}{16\pi G_2}\int d^2x \left(\,\frac{4r_0^2}{3L^2}\,\phi\,\partial_+\partial_-\Omega+\frac{1}{(x^+-x^-)^2}\Big(\frac{8r_0^2}{3L^2}\,\Omega\,\phi-16\phi^2\Big)\right)\ .
\end{equation}
Varying this action, we get the linearized equations of motion for the
perturbations,
\begin{eqnarray}\label{cbb2dleoms}
  \partial_+\partial_-\phi+\frac{2}{(x^+-x^-)^2}\,\phi&=&0\ ,
  \nonumber\\[3pt]
	\partial_+\partial_-\Omega+\frac{1}{(x^+-x^-)^2}\,\Big(2\,\Omega-\frac{24L^2}{r_0^2}\phi\Big)&=&0\ .
\end{eqnarray}
These equations are consistent at linear order with the ``constraint''
equations for the $++$ and $--$ components of the Einstein equation in
(\ref{cbb2deeoms}).
From these linearized equations, we see that the dilaton fluctuation
$\phi$ is decoupled from the metric fluctuation $\Omega$. Solving the
equation for $\phi$ in \eqref{cbb2dleoms}, we get
\begin{equation}\label{phipertRel}
	\phi=\frac{a+bt+c(t^2-\rho^2)}{\rho}\ ,
\end{equation}
where $a$, $b$, $c$ are independent constants.
Substituting the solution
(\ref{phipertRel}) for $\phi$ in the equation for $\Omega$ in
\eqref{cbb2dleoms}, we can solve for the metric perturbation $\Omega$,
which implies that the $AdS_2$ metric gets corrected at the same order
as the dilaton. The on-shell (boundary) action obtained then 
by using the linearized field equations in \eqref{cbb2dS2} gives terms
at quadratic order in the perturbations,
\begin{equation}\label{cbb2dS2bdy}\begin{split}
S_2=\frac{1}{16\pi G_2}&\int dt\sqrt{-\gamma}\,n^{\mu}\Big(\, \frac{2r_0^2}{3L^2}
    (\Omega\,\partial_{\mu}\phi-\phi\,\partial_{\mu}\Omega)\Big)\ ,
\end{split}\end{equation}
where $n^{\mu}$ is the outward unit normal to the boundary.

\subsubsection{The Schwarzian effective action}\label{Schw:rel}

In this section, we switch to Euclidean time $\tau=it$. The
Gibbons-Hawking boundary term in the 2-dimensional theory arises from
the reduction of the corresponding term in the higher dimensional
theory.  The Gibbons-Hawking term on the $3$-dimensional boundary of
the $4$-dimensional theories described by the Euclidean form of the action
\eqref{cbb4daction} is
\begin{equation}\label{4dGH}
	S_{GH}^{4d}=-\frac{1}{8\pi G_4}\int d^3x \sqrt{\gamma^{(3)}}\, K^{(4)}\ ,
\end{equation}
where the extrinsic curvature is defined as
$K^{(4)}_{AB}=\frac{1}{2}(\nabla_{A}n_{B}+\nabla_{B}n_{A})$, $n^{A}$
being the outward unit normal to the $3$-dimensional boundary. Using
the ansatz \eqref{compactmetric} for the $T^2$-compactification,
dimensionally reducing and performing the Weyl transformation of the
$2$-dimensional metric $g_{\mu\nu}=\Phi g^{(2)}_{\mu\nu}$, the
Gibbons-Hawking term reduces to\footnote{We
  have\ $K^{(4)}=\gamma^{(3)\,AB}K^{(4)}_{AB}=\gamma^{(3)\,\tau\tau}K^{(4)}_{\tau\tau}+2\gamma^{(3)\,xx}K^{(4)}_{xx}$,
  with
  $K^{(4)}_{xx}=-\Gamma^{r}_{xx}n_{r}=\frac{1}{2}n_{r}\partial^r\Phi^2=\frac{1}{2}n_{\mu}\partial^{\mu}\Phi^2$
  becomes $K^{(4)}=K^{(2)}+\Phi^{-2}n_{\mu}\partial^{\mu}\Phi^2$. Then
  \eqref{4dGH} gives \eqref{4dGHto2d} after the Weyl transformation.}
\begin{equation}\label{4dGHto2d}
S_{GH}^{4d}=-\frac{1}{16\pi G_2}\int d\tau\sqrt{\gamma}\,\Big(2\,\Phi^2K+\frac{3}{2}\,n_{\mu}\,\partial^{\mu}\Phi^2\Big)\ .
\end{equation}
The Ricci scalar term in the bulk 4-dim Euclidean action upon dimensional
reduction and after the Weyl transformation becomes
\begin{equation}\label{4driccireduction}
	-\sqrt{g^{(4)}}\mathcal{R}^{(4)}=-\sqrt{g}\Big(\Phi^2\mathcal{R}-\frac{3}{2}\nabla^2\Phi^2\Big)\ .
\end{equation}
Note also that $\sqrt{g^{(4)}}=\sqrt{g^{(2)}} \Phi^2$ and $\Phi^2=g_{xx}$\,.
We write the the total derivative term (the second term) in
\eqref{4driccireduction} as a boundary term
\begin{equation}
	-\frac{1}{16\pi G_2}\int d^2x\sqrt{g}\,\Big(-\frac{3}{2}\nabla^2\Phi^2\Big)=\frac{1}{16\pi G_2}\int d\tau\sqrt{\gamma}\,\Big(\frac{3}{2}\,n_{\mu}\partial^{\mu}\Phi^2\Big)\ .
\end{equation}
We see that this boundary term which comes from the dimensional
reduction of the bulk action in $4$-dimensions cancels the second term
in \eqref{4dGHto2d}, thereby giving the Gibbons-Hawking term on the
boundary of the $2$-dimensional theory as
\begin{equation}
	S_{GH}=-\frac{1}{8\pi G_2}\int d\tau\sqrt{\gamma}\, \Phi^2 K\ .
\end{equation}
Expanding the Gibbons-Hawking term in the perturbations
\eqref{cbb2dperturbations} and adding it to the Euclidean form of $S_2$ (which is $S_2^E=-iS_2$, with $t=-i\tau$ in $S_2$), the leading
term in the total boundary action $I_{bdy}=S_2^E+S_{GH}$ arises at
linear order in the dilaton perturbation (with subleading terms at
quadratic order). To illustrate this in greater
detail, it is important that we define the dilaton perturbation in
\eqref{cbb2dperturbations} in a physically appropriate manner. Since
the background value $\Phi_b$ is constant, it is sensible to define the
dilaton perturbation as
\be\label{dilpertz=1}
\Phi = \Phi_b\, (1+{\tilde\phi})\ ,\qquad \Phi_b = {r_0\over R}
\qquad \Rightarrow\qquad {\tilde\phi} = {\Phi-\Phi_b\over\Phi_b}\, \ll 1\ .
\ee
Thus with this redefinition, the perturbation is reasonable since it
automatically satisfies\ ${\tilde\phi}\ll 1$.\
In terms of the dilaton background value $\Phi_b$, the entropy
(\ref{entropyz=1}) is simply
\be\label{entropyz=1-2}
S_{BH}={\Phi_b^2\, V_2\over 4G_4} = {\Phi_b^2\over 4G_2}\ .
\ee
This gives
\begin{equation}\label{SchwEntrz=1}
  S_{GH}^{(1)}=-\frac{2\Phi_b^2}{8\pi G_2}\int d\tau\sqrt{\gamma}\
  {\tilde\phi}\, K\ \longrightarrow\
  -\frac{\Phi_b^2}{4\pi G_2}\int du\, \phi_r(u)\, \{\tau(u),u\}\ .
\end{equation}
In evaluating the last term, we take the boundary of $AdS_2$ as a
slightly deformed curve $(\tau(u),\rho(u))$ parametrized by the boundary
coordinate $u$, and define ${\tilde\phi}={\phi_r(u)\over\epsilon}$\,,\
as discussed in \cite{Maldacena:2016upp}\ (reviewed in
\cite{Sarosi:2017ykf}).\ Now using the outward unit normal $n^{\mu}$ to
the boundary, we expand the extrinsic curvature.
Expanding $S^{(1)}_{GH}$ then leads to a Schwarzian derivative action\
$Sch(\tau(u),u)=\{\tau(u),u\}={\tau'''\over \tau'}-{3\over 2} ({\tau''\over \tau'})^2$\,.\
The integral above pertains only to $AdS_2$ does not contain any further
scales besides the $AdS_2$ scale $L$ which also appears in the extrinsic
curvature giving the Schwarzian\ (also\ $\sqrt{\gamma}={L\over\epsilon}$).
The various length scales in the original extremal brane have been
absorbed into the $AdS_2$ scale $L$. Now we note that the coefficient
of the Schwarzian is in fact proportional to the entropy
(\ref{entropyz=1-2}) of the compactified extremal
black brane with $V_2$ finite\ (the dependence on $\Phi_b$ is expected
since it controls the transverse area). Since the entropy captures the
number of microstates of the unperturbed background, this is akin to a
central charge of the effective theory. Similar comments appear in 
\cite{Kitaev:2017awl}\ (see also \cite{Cvetic:2016eiv}, the
Schwarzian arising in some cases from the conformal anomaly).

It is worth noting that the coefficient in the Schwarzian term above
is proportional to the extremal entropy after the reasonable
definition of the perturbation as (\ref{dilpertz=1}) by scaling out
$\Phi_b$: apart from this, the Schwarzian term here is as in
\cite{Maldacena:2016upp}. As discussed there, we note that the
perturbation makes this nearly $AdS_2$ and contributes to the
near-extremal entropy via the Schwarzian. This can be obtained as in
the analysis there by a transformation\
$\tau(u)=\tan\frac{\tilde{\tau}(u)}{2}$\ which gives\
$S_{GH}^{(1)}=-{\Phi_b^2\over 4\pi G_2} {\bar\phi_r} \int du
\left(\{\tilde{\tau}(u),u\} + {1\over 2}\tilde{\tau}'^{\,2}\right)$\,,
treating ${\bar\phi_r}$ as constant.\ Solutions with
$\tilde{\tau}={2\pi\over\beta} u$ have
$\tilde{\tau}\sim\tilde{\tau}+2\pi$, giving the
action\ $S_{GH}^{(1)}=-2\pi^2 {\Phi_b^2\over 4\pi G_2} {\bar\phi_r}\,
T = -\log Z$, giving the near-extremal correction to the
entropy\ $\Delta S = 4\pi {\Phi_b^2\over 4 G_2} {\bar\phi_r}\,
T$\ (which, being linear in temperature, can also be seen to be the
specific heat): this again is proportional to the background entropy
with the perturbation defined as (\ref{dilpertz=1}).

The remaining terms in the expansion of $S_{GH}$ and $S^E_2$ are all
quadratic in perturbations and thus subleading compared to
$S_{GH}^{(1)}$. See also \eg\ \cite{Cvetic:2016eiv,Das:2017pif,
  Gaikwad:2018dfc,Nayak:2018qej}, for $AdS_2$ backgrounds
obtained from reductions of higher dimensional theories (see also
\cite{Cadoni:2017dma}). In particular there are parallels with some of
the analysis on the reduction of near extremal black holes in
\cite{Nayak:2018qej}.

Overall, expanding in the perturbations $\tilde{\phi}, \Omega$,
we have\ $I=S^E+S_{GH}=I_0+I_1+I_2+\dots$, with
\begin{equation}
  I_0=-\frac{\Phi_b^2}{16\pi G_2}
  \left(\int d^2x \sqrt{g}\,\mathcal{R}+2\int_{{\tiny bndry}}
  \sqrt{\gamma}\,K\right)
\end{equation}
is the background Euclidean action (see (\ref{S0z=1})): it can be checked that
$U(\Phi_b)=0$. The action $I_0$ is a topological term and gives the
extremal entropy $S_{BH}=\frac{\Phi_b^2}{4G_2}$\ after regulating this
as a near-extremal background\footnote{Here the Euclidean time
  periodicity, large for a small near-extremal temperature, precisely
  cancels the small regularized change in the extremal horizon. In
  more detail, expanding $f(r)$ in (\ref{cbb4dsolne}) about
  extremality, we have\ $f(r)\simeq {6 (r-r_0)\over r_0^2}
  (r-r_0+{r_0\over 6}(3-{Q^2\over r_0^4})) \equiv {6\over
    r_0^2}(r-r'_0-{\delta\over 2})(r-r'_0+{\delta\over
    2})$\ where\ $\delta={r_0\over 6}(3-{Q^2\over r_0^4})$ and
  $r'_0=r_0-\frac{\delta}{2}$.\ Then the nearly $AdS_2$ throat
  acquires a small horizon with metric $ds^2\sim {9\delta^2\over
    R^4}\rho^2d\tau^2+d\rho^2$ near the origin: the Euclidean time
  periodicity then is\ $\Delta\tau=\beta={2\pi R^2\over 3\delta}$
  consistent with (\ref{cbb4dextremalitye}).\ The horizon contribution
  to the action gives\ $I_0=-{\Phi_b^2\over 16\pi G_2}
  \Delta\tau\,{\delta\over 2} ({12\over R^2}) \equiv -\beta F$ and
  thereby the background extremal entropy $S_{BH}=-I_0$.\\ The
  boundary terms in the action above cancel: to elaborate, we have the
  $AdS_2$ metric $ds^2={L^2\over\rho^2}(d\tau^2+d\rho^2)$\,. The
  boundary at $\rho=\epsilon$ has outward unit normal
  $n_\rho=-{L\over\rho}$\,.  The extrinsic curvature defined
  as\ $K_{\mu\nu}={1\over 2} (\nabla_\mu n_\nu+\nabla_\nu
  n_\mu)$\ gives $K_{\tau\tau}=-\Gamma_{\tau\tau}^\rho
  n_\rho={L\over\rho^2}$\ and\ $K=\gamma^{\tau\tau}K_{\tau\tau}={1\over
    L}$.\ Then the terms at the boundary cancel as\ $-{\Phi_b^2\over
    16\pi G_2} (\int d\tau {L^2d\rho\over\rho^2}|^{hrzn}_\epsilon
  (-{2\over L^2}) + 2\int d\tau{L\over\epsilon} (-{1\over L}))$\,.}.
The linear terms are contained in
\begin{equation}
  I_1=-\frac{2\Phi_b^2}{16\pi G_2}\int d^2x\,\sqrt{g}\,
  \tilde{\phi}\,\Big(\mathcal{R}-\frac{\partial U}{\partial\Phi^2}\Big)
  -\frac{2\Phi_b^2}{8\pi G_2}\int_{bndry} \sqrt{\gamma}\,\tilde{\phi}\,K\ ,
\end{equation}
with $\frac{\partial U}{\partial\Phi^2}|_{\Phi_b}=-\frac{2}{L^2}$,
which is the Jackiw-Teitelboim theory
\cite{Jackiw:1984je,Teitelboim:1983ux}, which serves as a simple model
for $AdS_2$ physics (with parallels with the SYK model). The bulk term
vanishes by the $\tilde{\phi}$ equation giving the fixed background
$AdS_2$ geometry, while the boundary term gives the Schwarzian as
explained above. The analysis here of the higher dimensional
realization serves to recover the background entropy as expected and
reveal the various subleading terms beyond the Jackiw-Teitelboim
theory emerging from reduction: $I_2$ is second order in
perturbations, from $S^E_2$ (see (\ref{cbb2dS2bdy})) and the second
order terms in the expansion of $S_{GH}$,
\be\label{I2z=1}
I_2=-\frac{1}{16\pi G_2}\int d\tau\sqrt{\gamma}\Big[\, \frac{2r_0^2}{3L^2}\,
\Phi_b n^{\rho} (\Omega\,\partial_{\rho}{\tilde\phi}-{\tilde\phi}\,\partial_{\rho}\Omega)+2\Phi_b^2(\tilde{\phi}^2\,K-2\tilde{\phi}e^{-\omega_b}\partial_{\rho}\Omega)\Big]\ ,
\ee
expanding in conformal gauge.

\section{Charged hyperscaling violating Lifshitz black branes}\label{sec:chvLif}

Over the last several years, nonrelativistic generalizations of
holography have been investigated extensively: see \eg\
\cite{Hartnoll:2016apf} for a review of various aspects. A
particular family of interesting theories comprises the so-called
hyperscaling violating Lifshitz (hvLif) theories, which are conformal
to Lifshitz theories. These arise as solutions to
Einstein-Maxwell-scalar theories, the $U(1)$ gauge field and dilaton
scalar necessary to support the nonrelativistic background. For the
most part, we regard these as effective gravity theories: in certain
cases these can be shown to arise from gauge/string realizations (see
\eg\ \cite{Narayan:2012hk}).

These nonrelativistic black branes are uncharged. A minimal way to
construct charged black branes is to add an additional $U(1)$ gauge
field, which serves to supply charge to the black brane: see
\eg\ \cite{Tarrio:2011de}, \cite{Alishahiha:2012qu},
\cite{Bueno:2012sd}. For these latter charged black branes, there
exist extremal limits where the near horizon geometry takes the
form\ $AdS_2\times X$, and contains an $AdS_2$ throat.  Compactifying
the transverse space now allows us to study the extremal limits of
these theories in the context of a 2-dimensional dilaton gravity
theory with additional matter, notably the scalar descending from
higher dimensions as well as gauge fields\footnote{Note that in the
  $AdS/CMT$ literature, these theories are referred to
  Einstein-Maxwell-dilaton theories: we here use
  Einstein-Maxwell-scalar since the 2-dim dilaton $\Phi$ here is
  distinct from the hvLif scalar $\Psi$.}.

\subsection{$4$-dimensional charged hvLif black brane}

Consider Einstein-Maxwell-scalar theory with a further $U(1)$ gauge
field, with action \cite{Alishahiha:2012qu}
\begin{equation}\label{chbb4daction}\begin{split}
  S=\int d^4x\sqrt{-g^{(4)}}\Big[\frac{1}{16\pi G_4}\Big(\mathcal{R}^{(4)}-\frac{1}{2}\partial_M\Psi\partial^M\Psi+V(\Psi)-\frac{Z_1}{4}F_{1\,MN}F_1^{MN}\Big)-\frac{Z_2}{4}F_{2\,MN}F_2^{MN}\Big],
\end{split}\end{equation}
where the scalar field dependent couplings and the scalar potential are
\begin{equation}\label{chbbcouplings}
	Z_1=e^{\lambda_1\Psi}\ , \qquad Z_2=e^{\lambda_2\Psi}\ , \qquad V(\Psi)=V_0 e^{\gamma\Psi}\ .
\end{equation}
The field equations following from the above action are
\bea\label{chbb4dEOM}
&& \mathcal{R}^{(4)}_{MN}-\frac{1}{2}\partial_{M}\Psi\partial_{N}\Psi+g_{MN}\frac{V}{2}-\frac{Z_1}{2}\Big(F_{1\,MP}F_{1N}^{\ \ P}-\frac{g_{MN}}{4}(F_1)^2\Big)
\nonumber\\
&&\hspace{60mm} -8\pi G_4Z_2\Big(F_{2\,MP}F_{2N}^{\ \ P}-\frac{g_{MN}}{4}(F_2)^2\Big)=0 ,\nonumber\\[2pt]
&&	\frac{1}{\sqrt{-g^{(4)}}}\partial_{M}(\sqrt{-g^{(4)}}\partial^{M}\Psi)+\gamma V-\frac{\lambda_1 Z_1}{4}F_{1\,MN}F_1^{MN}-4\pi G_4\lambda_2 Z_2 F_{2\,MN}F_2^{MN}=0\ , \nonumber\\[2pt]
&& \partial_{M}(\sqrt{-g^{(4)}}Z_1F_1^{MN})=0\ , \qquad \partial_{M}(\sqrt{-g^{(4)}}Z_2F_2^{MN})=0\ .
\eea
The charged hvLif black brane solution to these equations is
\begin{eqnarray}
	ds^2&=&\Big(\frac{r}{r_{hv}}\Big)^{-\theta}\Big[-\frac{r^{2z}f(r)}{R^{2z}}dt^2+\frac{R^2}{r^2f(r)}dr^2+\frac{r^2}{R^2}(dx^2+dy^2)\Big]\ , \nonumber\\[2pt]
	f(r)&=&1-\left(\frac{r_0}{r}\right)^{2+z-\theta}+\frac{Q^2}{r^{2(1+z-\theta)}}\Big(1-\Big(\frac{r}{r_0}\Big)^{z-\theta}\Big)\ , \nonumber\\[4pt]
F_{1rt}&=&\sqrt{2(z-1)(2+z-\theta)}\,e^{-\frac{\lambda_1\Psi_0}{2}}\,r_{hv}^2\,
R^{\theta-z-4}\, r^{1+z-\theta}\ , \nonumber\\[4pt]
F_{2rt}&=&\frac{Q\sqrt{2(2-\theta)(z-\theta)}\,e^{-\frac{\lambda_2\Psi_0}{2}}}{4\sqrt{\pi G_4}}\,R^{z-\theta-2}\,r_{hv}^{-z+\theta+1}\,r^{-(1+z-\theta)}\ ,
\nonumber\\[4pt]
	e^{\Psi}&=&e^{\Psi_0}\Big(\frac{r_{hv}\,r}{R^2}\Big)^{\sqrt{(2-\theta)(2z-2-\theta)}}\ , \label{chbb4dsoln}
\end{eqnarray}
being explicit with length scales, and
\begin{equation}\label{chbb4dparameters}\begin{split}
	& V_0=\frac{(2+z-\theta)(1+z-\theta)\,e^{-\gamma\Psi_0}}{R^{2-2\theta}\,r_{hv}^{2\theta}}\ , \qquad \gamma=\frac{\theta}{\sqrt{(2-\theta)(2z-2-\theta)}}\ ,\\[2pt]
	& \lambda_1=\frac{-4+\theta}{\sqrt{(2-\theta)(2z-2-\theta)}}\ , \qquad \lambda_2=\sqrt{\frac{2z-2-\theta}{2-\theta}}\ .
\end{split}\end{equation}
Here $r_{hv}$ is the hyperscaling violating scale arising in the conformal
factor in the metric, and the charge parameter $Q$ has dimensions of
$r^{1+z-\theta}$: this is equivalent to absorbing factors of $r_{hv}, R$
into $Q$.\ For $z=1, \theta=0$, this scaling coincides with that
for the relativistic black brane in sec.~\ref{sec:relElecbrane}.

In these charged hyperscaling violating Lifshitz black brane solutions
to the action \eqref{chbb4daction}, the gauge field $A_1$ and the
scalar field $\Psi$ source the hyperscaling violating Lifshitz
background while the gauge field $A_2$ giving charge to the black
brane, as mentioned above. This action (\ref{chbb4daction}) has also
been defined by absorbing the Newton constant into the definition of
the hyperscaling violating gauge field $A_1$ and scalar $\Psi$ (which
thus makes $A_1$ and $\Psi$ dimensionless) while retaining the gauge
field $A_2$ in $F_2$ as having mass dimension one. Thus the field strength
$F_{2\,rt}$ in (\ref{chbb4dsoln}) has mass dimension 2, as for the
relativistic brane.

The null energy conditions for the metric follow from the asymptotic
hvLif geometry \cite{Hartnoll:2016apf} and are given by
\begin{equation}\label{chbbnec}
  (z-1)(2+z-\theta)\geq 0\ , \qquad\quad
  (2-\theta)(2(z-1)-\theta)\geq 0\ .
\end{equation}
In addition, we require the gauge field $A_{2\,t}$ to vanish at the
boundary ($r\rightarrow\infty$) so that the theory does not ruin the
hvLif boundary conditions we have assumed: this is equivalent to
assuming that these charged black branes represent finite temperature
charged states in the boundary hvLif theory. The background profile
$A_{2\,t}\sim 1-({r_0\over r})^{z-\theta}$ then implies that
\begin{equation}\label{z>theta}
	z-\theta \geq 0\ .
\end{equation}
These conditions together constrain the range of $z$, $\theta$ for these
extremal nonrelativistic black brane backgrounds, which will be important
in the discussion of perturbations later.\ Specifically:\\
(i) First, the last condition (\ref{z>theta}) is specific to the charged
case: using this, the first of the null energy conditions (\ref{chbbnec})
implies that\ $z\geq 1$.\\
(ii) From the second of the conditions (\ref{chbbnec}), we have either
$2-\theta\geq 0,\ 2z-2-\theta\geq 0$,\ or\ 
$2-\theta < 0,\ 2z-2-\theta < 0$.\ Considering the second possibility,
we obtain\ $z\geq \theta\geq 2$, but this implies\ $2z-2-\theta=z-2+z-\theta>0$,
which is a contradiction.\ This forces\ $2-\theta\geq 0,\ 2z-2-\theta\geq 0$.\\
Overall, this gives the conditions
\be\label{chbbzthetarange}
z\geq 1\ ,\qquad\quad 2z-2-\theta\geq 0\ ,\qquad\quad 2-\theta\geq 0\ ,
\ee
for the regime of validity of the $z,\theta$ exponents of the charged
hvLif background above. For the special case of $z=1$, the NEC becomes
$(2-\theta)(-\theta)\geq 0$, which forces $\theta\leq 0$ by
(\ref{chbbzthetarange}).

The relativistic limit of this charged hvLif black brane gives the
relativistic electric black brane discussed previously in
sec.~\ref{sec:relElecbrane}. From the constraint \eqref{chbbzthetarange},
we see that the correct relativistic limit is to take first $\theta=0$
and then $z=1$. In this limit, we get
\begin{equation}
	\gamma=0, \qquad \lambda_1\rightarrow-\infty, \qquad \lambda_2=0, \qquad V_0=6/R^2, \qquad \Psi=\Psi_0.
\end{equation}
With this the Einstein-Maxwell-scalar action \eqref{chbb4daction}
reduces to the Einstein-Maxwell action \eqref{cbb4daction}, where
$F_2$ and $V_0$ in \eqref{chbb4daction} are identified with $F$ and
$-2\Lambda$ in \eqref{cbb4daction}.

\subsubsection{Extremality and attractors}

In the extremal limit,
\begin{equation}
  T=\frac{(2+z-\theta)r_0^z}{4\pi R^{z+1}}\Big(1-\frac{(z-\theta)Q^2 r_0^{-2(1+z-\theta)}}{(2+z-\theta)}\Big)=0 \ \ \implies \
  Q^2=\frac{(2+z-\theta)}{(z-\theta)}r_0^{2(1+z-\theta)},\ 
\end{equation}
and the near horizon geometry becomes $AdS_2\times \mathbb{R}^2$,
\begin{equation}\begin{split}
		ds^2&=\Big(\frac{r_0}{r_{hv}}\Big)^{-\theta}\Big[-\frac{r_0^{2z}f(r)}{R^{2z}}dt^2+\frac{R^2}{r_0^2f(r)}dr^2+\frac{r_0^2}{R^2}(dx^2+dy^2)\Big]\ ,\\[2pt]
	& f(r)|_{r\rightarrow r_0}\simeq \frac{(2+z-\theta)(1+z-\theta)}{r_0^2}(r-r_0)^2\ ,
\end{split}\end{equation}
the $AdS_2$ scale being $R\, ({r_0\over r_{hv}})^{-\theta/2}$.\ 
The Bekenstein-Hawking entropy is the horizon area in Planck units\
\be\label{entropyztheta}
S_{BH} = \Big({r_0^2\over R^2}\Big)\, \Big({r_0\over r_{hv}}\Big)^{-\theta}\,
{V_2\over 4G_4}\ =\
\Big({z-\theta\over 2+z-\theta}\Big)^{{2-\theta\over 2(1+z-\theta)}}\
    {r_{hv}^\theta\, V_2\over 4G_4}\, {Q^{(2-\theta)/(1+z-\theta)}\over R^2}\ ,
\ee
where $V_2=\int dx dy$ is the transverse area of the brane.
For $z=1, \theta=0$, this coincides with the relativistic brane.

It is worth noting that the full metric in (\ref{chbb4dsoln}) is
asymptotically of hvLif form, for $r\gg r_0$. The boundary of the theory
could be taken as $r\sim r_{hv}$, \ie\ the theory flows to hvLif below
this scale, in some bigger phase diagram. The $AdS_2$ throat,
well-defined if\ ${r-r_0\over r_0}\ll 1$ and ${r-r_0\over R}\gg 1$, 
is well-separated from the asymptotic hvLif region if
${r-r_0\over r_{hv}}\ll 1$\, and the $AdS_2$ scale satisfies\
$R\, ({r_0\over r_{hv}})^{-\theta/2} \ll r_{hv}$\,\ \ie\
$R\ll r_{hv} ({r_0\over r_{hv}})^{\theta/2}$\,.\ Note that this is
not vacuous since $r_0\ll r_{hv}$ so that ${r_0\over r_{hv}}\ll 1$ is a
small factor.

Along the lines of the attractor mechanism discussion in
\cite{Goldstein:2005hq}, we would like to convert this theory to a
dilatonic gravity theory in 4-dimensions with a potential (and no gauge
fields). Towards this end, we integrate Maxwell's equations in
\eqref{chbb4dEOM} and use the solutions for field strengths in
\eqref{chbb4dsoln} to get
\begin{equation}\label{chbb4dF1F2}
F_1^{tr}=\frac{\sqrt{2(z-1)(2+z-\theta)}\;e^{\frac{\lambda_1\Psi_0}{2}}\,
  r_{hv}^{\theta-2}\,R^{1-\theta}}{\sqrt{-g}\;e^{\lambda_1\Psi}}\ , \qquad
F_2^{tr}=\frac{Q\sqrt{2(2-\theta)(z-\theta)}\;e^{\frac{\lambda_2\Psi_0}{2}}\,
  r_{hv}^{z-1}}{4\sqrt{\pi G_4}\,R^{2z+1-\theta}\;\sqrt{-g}\;e^{\lambda_2\Psi}}\ .
\end{equation}
Substituting \eqref{chbb4dF1F2} in \eqref{chbb4dEOM}, we obtain equations
of motion for the metric and the scalar field $\Psi$, which can be
derived from the following equivalent action
\bea\label{VeffPsiAction}
&&  S = \frac{1}{16\pi G_4}\int d^4 x\sqrt{-g}\Big(\mathcal{R}-\frac{1}{2}(\partial\Psi)^2-V_{eff}(\Psi)\Big)\ , \nonumber\\[2pt]
&& V_{eff}(\Psi)=-\frac{(2+z-\theta)(1+z-\theta)}{R^{2-2\theta}r_{hv}^{2\theta}}e^{\gamma(\Psi-\Psi_0)} \\
&& \qquad\qquad\ \
+ \frac{1}{g_{xx}^2}\Big(\frac{(z-1)(2+z-\theta)r_{hv}^{2\theta-4}R^{2-2\theta}}{e^{\lambda_1(\Psi-\Psi_0)}}+\frac{(2-\theta)(z-\theta)Q^2 r_{hv}^{2z-2}R^{-4z-2+2\theta}}{e^{\lambda_2(\Psi-\Psi_0)}}\Big)\ .\nonumber
\eea
The explicit scales show that the potential term-by-term has mass
dimension 2.
This equivalent action is obtained by substituting the solutions for
$F_1^{tr}$ and $F_2^{tr}$ in the action \eqref{chbb4daction} and
changing the signs of $F_1^2$, $F_2^2$ terms, as earlier.
At the critical point (extremality),
\begin{equation}\label{chbb4dexvalues}
	g_{xx}=\Big(\frac{r_0}{r_{hv}}\Big)^{-\theta}\Big(\frac{r_0}{R}\Big)^2, \qquad e^{\Psi}=e^{\Psi_0}\Big(\frac{r_{hv}\,r_0}{R^2}\Big)^{\sqrt{(2-\theta)(2z-2-\theta)}}, \qquad Q^2=\frac{(2+z-\theta)}{(z-\theta)}r_0^{2(1+z-\theta)},
\end{equation}
the first and second derivatives of $V_{eff}$ (\eqref{chbb4dVeff'}, \eqref{chbb4dVeff''}) are
\begin{equation}\label{VeffPsiStab}
  \frac{\partial V_{eff}}{\partial\Psi}\Big|_{ext}=0\ , \qquad\quad
  \frac{\partial^2 V_{eff}}{\partial\Psi^2}\Big|_{ext}=\frac{4(z-1)(2+z-\theta)(1+z-\theta)}{2z-2-\theta}\frac{r_0^{\theta}}{r_{hv}^{\theta}R^2}\,>\,0\ ,
\end{equation}
which imply that the extremal point is stable for all values of $z$,
$\theta$ allowed by the conditions \eqref{chbbzthetarange}.\ \ It is
worth mentioning that for $z=1$ and $\theta$ nonzero, these and all higher
derivatives of $V_{eff}$ in fact vanish (see (\ref{chbb4dVeff'n})): thus
we obtain no insight into the stability of these attractors in this
case and we will not discuss this subcase in what follows.

\subsection{Dimensional reduction to $2$-dimensions}

Compactifying the two spatial dimensions, $x^i$ as $T^2$, we 
dimensionally reduce with the metric ansatz (\ref{compactmetric}), taking
the lower dimensional fields $g^{(2)}_{\mu\nu}, \Phi, \Psi, A_1, A_2$,
to be $T^2$-independent: then 
the action \eqref{chbb4daction} reduces to \eqref{chbb2daction}.
Performing a Weyl transformation, $g_{\mu\nu}=\Phi g^{(2)}_{\mu\nu}$
to absorb the kinetic term for the dilaton $\Phi$ in the Ricci scalar,
the $2$-dimensional action \eqref{chbb2daction} becomes
\begin{equation}\label{chbb2dactionWeyl}
  S=\int d^2 x\sqrt{-g}\Big[\frac{1}{16\pi G_2}\Big(\Phi^2\mathcal{R}-\frac{\Phi^2}{2}\partial_{\mu}\Psi\partial^{\mu}\Psi+V\Phi-\frac{\Phi^3}{4}Z_1 F_{1\,\mu\nu}F_1^{\mu\nu}\Big)-\frac{V_2\Phi^3}{4}Z_2 F_{2\,\mu\nu}F_2^{\mu\nu}\Big]\ .
\end{equation}
We only retain fields with nontrivial background profiles: more general
comments appear later. The Maxwell equations for the gauge fields are
\begin{equation}\label{chbb2dF1F2EWeyl}
  \partial_{\mu}(\sqrt{-g}\Phi^3 Z_1F_1^{\mu\nu})=0\ , \qquad
  \partial_{\mu}(\sqrt{-g}\Phi^3 Z_2F_2^{\mu\nu})=0\ .
\end{equation}
Integrating and using $F_{1rt}, F_{2rt}$ from (\ref{chbb4dsoln}) to fix
the integration constants gives
\begin{equation}\label{chbb2dF1F2soln}
F_1^{\mu\nu}=\frac{\sqrt{2(z-1)(2+z-\theta)}\,e^{\frac{\lambda_1\Psi_0}{2}}\,
r_{hv}^{\theta-2}\,R^{1-\theta}}{\sqrt{-g}\;Z_1\;\Phi^3}\,\varepsilon^{\mu\nu},
\quad
F_2^{\mu\nu}=\frac{Q\sqrt{2(2-\theta)(z-\theta)}\,e^{\frac{\lambda_2\Psi_0}{2}}\,
r_{hv}^{z-1}}{4\sqrt{\pi G_4}\,R^{2z+1-\theta}\,\sqrt{-g}\;Z_2\,\Phi^3}\,\varepsilon^{\mu\nu},
\end{equation}
where $\varepsilon^{\mu\nu}$ satisfies\ $\varepsilon^{tr}=1=-\varepsilon^{rt}$
and $\varepsilon_{\mu\nu}=g_{\mu\rho}g_{\nu\sigma}\varepsilon^{\rho\sigma}$. We
substitute the solutions \eqref{chbb2dF1F2soln} in the remaining field
equations obtained by varying the action \eqref{chbb2dactionWeyl} (\ie\
eq.~\eqref{chbb2dnoWeyleoms}) to obtain
\bea\label{chbb2dequiveoms}  
g_{\mu\nu}\nabla^2\Phi^2-\nabla_{\mu}\nabla_{\nu}\Phi^2
  +\frac{g_{\mu\nu}}{2}\Big(\frac{\Phi^2}{2}(\partial\Psi)^2+U\Big)
  -\frac{\Phi^2}{2}\partial_{\mu}\Psi\partial_{\nu}\Psi &=& 0\ ,\nonumber\\
\mathcal{R}-\frac{1}{2}(\partial\Psi)^2 -
\frac{\partial U}{\partial(\Phi^2)} &=& 0\ ,\nonumber\\
\frac{1}{\sqrt{-g}}\partial_{\mu}(\sqrt{-g}\Phi^2\partial^{\mu}\Psi)
  -\frac{\partial U}{\partial\Psi} &=& 0\ ,
\eea
where $U(\Phi,\Psi)$ is an effective interaction potential. These
equations can then be obtained from the following equivalent action
\bea\label{chbb2deffactionWeyl}
&& S=\frac{1}{16\pi G_2}\int d^2 x\sqrt{-g}\ \Big(\Phi^2\mathcal{R}-\frac{\Phi^2}{2}(\partial\Psi)^2-U(\Phi,\Psi)\Big), \\[2pt]
&& U(\Phi,\Psi)=-\frac{(2+z-\theta)(1+z-\theta)}{R^{2-2\theta}r_{hv}^{2\theta}}e^{\gamma(\Psi-\Psi_0)}\,\Phi \nonumber\\
&& \quad +\frac{1}{\Phi^3}\Big(\frac{(z-1)(2+z-\theta)r_{hv}^{2\theta-4}R^{2-2\theta}}{e^{\lambda_1(\Psi-\Psi_0)}}+\frac{(2-\theta)(z-\theta)Q^2 r_{hv}^{2z-2}R^{-4z-2+2\theta}}{e^{\lambda_2(\Psi-\Psi_0)}}\Big)\ , \nonumber
\eea
where $V_0$, $\gamma$, $\lambda_1$, $\lambda_2$ are given in \eqref{chbb4dparameters}. This equivalent action is
obtained by substituting the solutions for $F_1^{\mu\nu}$,
$F_2^{\mu\nu}$ in terms of the dilaton $\Phi^2$ and the scalar $\Psi$
in the action \eqref{chbb2dactionWeyl} and changing the signs of
$F_1^2$, $F_2^2$ terms, as discussed in the case for relativistic
electric black brane, sec.~\ref{sec:relElecbrane}. Also note that the
relativistic electric black brane is a special case of the
dilaton-gravity-matter theory, considered here, for $\theta=0$ and $z=1$.

We note that the scalar $\Psi$ that descends from the hyperscaling
violating scalar in higher dimensions is not minimally coupled in the
2-dimensional theory. The potential $U(\Phi,\Psi)$ contains nontrivial
interactions between the dilaton $\Phi$ and the hvLif scalar
$\Psi$. Thus the small fluctuation spectrum of the dilaton and $\Psi$
are coupled, and one might worry about the stability of the
2-dimensional attractor. This is reminiscent of multi-field inflation
models, where one scalar field provides a slow-roll phase while another
scalar provides a waterfall phase, ending inflation. In the current
context, stability would require that no tachyonic modes arise from
the interaction induced by $U(\Phi,\Psi)$ between $\Phi$ and $\Psi$. We
will address this soon.

The field equations (\ref{chbb2dequiveoms}) admit a constant dilaton,
$AdS_2$ solution as
\bea\label{chbb2dexads2}
ds^2= L^2 \Big[-{r_0^{2z-3\theta}\over R^{2z}r_{hv}^{-3\theta} L^4}\,(r-r_0)^2 dt^2
+ \frac{dr^2}{(r-r_0)^2}\Big] , &&
L^2\equiv \frac{Rr_0^{1-\frac{3\theta}{2}}r_{hv}^{\frac{3\theta}{2}}}{(2+z-\theta)(1+z-\theta)}\ ,\nonumber\\
\medspace
\Phi^2=\Big(\frac{r_0}{r_{hv}}\Big)^{-\theta}\Big(\frac{r_0}{R}\Big)^2, &&
e^{\Psi}=e^{\Psi_0}\Big(\frac{r_{hv}\,r_0}{R^2}\Big)^{\sqrt{(2-\theta)(2z-2-\theta)}},\nonumber\\
Q^2=\frac{(2+z-\theta)}{(z-\theta)}r_0^{2(1+z-\theta)}. &&
\eea
Let us choose conformal gauge by doing a coordinate transformation,
\begin{equation}
\rho=\frac{R^{z+1}r_0^{1-z}}{(2+z-\theta)(1+z-\theta)}\frac{1}{(r-r_0)}\ .
\end{equation}
In conformal gauge, the $AdS_2$ metric in \eqref{chbb2dexads2} can be written as
\begin{equation}\label{chbb2dexads2cg}\begin{split}
ds^2&=e^{2\omega}(-dt^2+d\rho^2)=e^{2\omega}(-dx^+dx^-), \qquad e^{2\omega}=\frac{L^2}{\rho^2}\ ,
\end{split}\end{equation}
where the lightcone coordinates are $x^{\pm}=t\pm\rho$ and $L$ is the radius of $AdS_2$. To see that \eqref{chbb2deffactionWeyl} admits the above $AdS_2$ solution, we compute $\frac{\partial U}{\partial\Phi^2}$ for the above solution, which gives
\begin{equation}
	\frac{\partial U}{\partial\Phi^2}=-2\frac{(2+z-\theta)(1+z-\theta)}{Rr_0^{1-\frac{3\theta}{2}}r_{hv}^{\frac{3\theta}{2}}}=-\frac{2}{L^2}\ .
\end{equation}
From \eqref{chbb2dequiveoms} for $\Psi=constant$ (from
\eqref{chbb2dexads2}), we get the Ricci scalar as
\begin{equation}
	\mathcal{R}=\frac{\partial U}{\partial\Phi^2}=-\frac{2}{L^2}\ .
\end{equation}

\subsubsection{Perturbations about $AdS_2$}

As before, we turn on perturbations to fields with background profiles, 
\ie\ to the metric, the dilaton $\Phi$ and the scalar field $\Psi$,
\begin{equation}\label{chbb2dperturbations}
	\Phi=\Phi_b + \phi(x^+,x^-)\ , \quad\ \ \omega=\omega_b+\Omega(x^+,x^-)\ , \quad\ \  \Psi=\Psi_b+\sqrt{2z-2-\theta}\;\psi(x^+,x^-)\ ,\ \
\end{equation}
where $\Phi_b$, $\omega_b$ and $\Psi_b$ denote the
\eqref{chbb2dexads2} background solution. Expanding the action
\eqref{chbb2deffactionWeyl} (in conformal gauge) about this background gives
\be
S = \frac{1}{16\pi G_2}\int d^2x\ \Big(4\Phi^2\partial_+\partial_-\omega+\Phi^2\partial_+\Psi\partial_-\Psi-\frac{e^{2\omega}}{2}U(\Phi,\Psi)\Big)
\ \equiv\ S_0+S_1+S_2\ ,
\ee
where
\begin{equation}\label{S0ztheta}
	S_0=\frac{1}{16\pi G_2}\int d^2x\ \Big(4\Phi_b^2\partial_+\partial_-\omega_b+\Phi_b^2\partial_+\Psi_b\partial_-\Psi_b-\frac{e^{2\omega_b}}{2}U(\Phi_b,\Psi_b)\Big)
\end{equation}
is the background action and $S_1$ vanishes by the equations of motion.
$S_2$ is quadratic in perturbations and is given by
\begin{equation}\label{actionPertS2}
  \begin{split}
    S_2=&\frac{1}{16\pi G_2}\int d^2x\, \frac{r_0^{2-2\theta}r_{hv}^{2\theta}}{L^2(2+z-\theta)(1+z-\theta)}\ \Big[8\,\phi\,\partial_+\partial_-\Omega+\frac{16}{(x^+-x^-)^2}\phi\,\Omega\\
    &\ \ +\frac{r_0^{2-2\theta}r_{hv}^{2\theta}}{L^2(2+z-\theta)(1+z-\theta)}\Big((2z-2-\theta)\partial_+\psi\partial_-\psi-\frac{4(z-1)}{(x^+-x^-)^2}\psi^2\Big)\\
&\ \ +\frac{1}{(x^+-x^-)^2}\Big( -\frac{16L^2(2+z-\theta)(1+z-\theta)}{r_0^{2-2\theta}r_{hv}^{2\theta}}\phi^2+\frac{8\,\theta}{\sqrt{(2-\theta)}}\psi\phi\Big)\Big].
\end{split}\end{equation}
Varying this action, we get the linearized equations of motion for the
perturbations,
\begin{equation}
 \partial_+\partial_-\phi+\frac{2}{(x^+-x^-)^2}\,\phi = 0\ ,
\label{chbb2dlphieqn1} \nonumber
\end{equation}
\begin{equation}
 (2z-2-\theta)\partial_+\partial_-\psi+\frac{1}{(x^+-x^-)^2}\Big(4(z-1)\psi
-\frac{L^2(2+z-\theta)(1+z-\theta)}{r_0^{2-2\theta}r_{hv}^{2\theta}\;}\frac{4\theta}{\sqrt{(2-\theta)}}\phi\Big) = 0\ ,
\label{chbb2dlpsieqn}
\end{equation}
\medspace
\begin{equation}
 \partial_+\partial_-\Omega+\frac{1}{(x^+-x^-)^2}\left(2\,\Omega-\frac{4L^2(2+z-\theta)(1+z-\theta)}{r_0^{2-2\theta}r_{hv}^{2\theta}}\phi+\frac{\theta}{\sqrt{(2-\theta)}}\psi\right) = 0\ .
\label{chbb2dlOmegaeqn} \nonumber
\end{equation}
These equations are consistent at linear order with the ``constraint''
equations for the $\pm\pm$ components of the Einstein equation in
(\ref{chbb2dequiveoms}): see Appendix, eq.(\ref{chbb2deomsclc})-(\ref{chbb2deomsclcconstraint2}).\\
We see that the equation for $\psi$ is coupled to $\phi$ as well: 
defining a new field $\zeta$,
\begin{equation}
  \zeta=\psi-\frac{2}{\sqrt{2-\theta}}\,
  \frac{L^2(2+z-\theta)(1+z-\theta)}{r_0^{2-2\theta}r_{hv}^{2\theta}}\, \phi\ ,
\end{equation}
decouples the equations for $\zeta$ and $\phi$, which now become
\begin{equation}\label{chbb2dldeoms}\begin{split}
	\partial_+\partial_-\phi+\frac{2}{(x^+-x^-)^2}\,\phi=&0\ ,\\
	(2z-2-\theta)\partial_+\partial_-\zeta+2(z-1){2\over (x^+-x^-)^2}\,\zeta=&0\ , \\
	\partial_+\partial_-\Omega+\frac{1}{(x^+-x^-)^2}\Big(2\,\Omega+\frac{2(3\theta-4)}{(2-\theta)}\frac{L^2(2+z-\theta)(1+z-\theta)}{r_0^{2-2\theta}r_{hv}^{2\theta}}\,\phi+\frac{\theta}{\sqrt{(2-\theta)}}\,\zeta\Big)=&0.
\end{split}\end{equation}
In this form, the perturbations $\phi$ and $\zeta$ are equivalent to scalars
with positive mass propagating in a perturbed $AdS_2$ background, with
equation of motion\
${1\over\sqrt{-g}}\del_\mu(\sqrt{-g}g^{\mu\nu}\del_\nu\phi)-m^2\phi=0$:
in conformal gauge this is $\del_+\del_-\phi+{m^2L^2\over (x^+-x^-)^2}\phi=0$.\
Let us look at a few special cases here:
\begin{itemize}
\item{ For $z=1,\ \theta=0$, we have seen
that this system reduces to the relativistic brane case studied earlier
\eqref{cbb2dleoms}, and the $\Psi$ scalar (the
nonrelativistic scalar in higher dimensions) can be then seen to
decouple from the system: in particular, the terms containing
$\psi$-perturbations vanish in the action (\ref{actionPertS2}) for
quadratic perturbations. This is expected from the fact that the
original action for the higher dimensional nonrelativistic theory
reduces to the relativistic brane theory as $z\ra 1,\ \theta\ra 0$, as
discussed after (\ref{chbb4daction}). In effect, we have defined the
$\psi$-perturbation in (\ref{chbb2dperturbations}) so that the
relativistic brane limit arises smoothly, and the $\Psi$-scalar freezes
out. This is also reflected in the linearized equations for
perturbations.}
\item{ For $\theta=0$ and $z>1$, both $\phi$ and $\zeta$ have positive
mass term coefficients, and further $\zeta$ decouples entirely from the
$\Omega$ equation. This means that in fact any linear combination
of the fields $A\phi+B\zeta$ also in fact has a positive mass term
coefficient in its linearized fluctuation equation, as can be seen by
taking that linear combination of the two equations\
$\del_+\del_-(A\phi+B\zeta)+ {2\over (x^+-x^-)^2} (A\phi+B\zeta)=0$.
The linear fluctuation analysis thus suggests that the attractor point
is in fact perfectly stable for small fluctuations.}
\item{ For $\theta\neq 0$ and $z=1$, we see that the $\zeta$ field is a
massless mode and further it does not decouple from the $\Omega$ equation.
This suggests that the linear stability analysis is insufficient to
determine stability of the attractor point. However in this case, there
is a more basic concern: looking back at the higher dimensional system
(\ref{VeffPsiStab}), we see that in fact ${\del^2V_{eff}\over\del\Psi^2}=0$
in this case (in fact all derivatives vanish, (\ref{chbb4dVeff'n})), so that
the higher dimensional theory is also not manifestly
a stable attractor. Thus the relevance of the 2-dimensional theory is
less clear in this case.}
\item{ For generic $z,\theta$ values satisfying the energy conditions
(\ref{chbbnec}), (\ref{z>theta}), (\ref{chbbzthetarange}), we see that
the mass term coefficients for both $\phi$ and $\zeta$ perturbations are
positive. Now a generic linear combination of the fields $A\phi+B\zeta$
satisfies\
\be
\del_+\del_-\left(A\phi+(2z-2-\theta)B\zeta\right)+ {2\over (x^+-x^-)^2}
\left(A\phi+(2z-2-\theta)B\zeta\right) =
- {2\over  (x^+-x^-)^2} \theta\,B\zeta\ .
\ee
This is akin to a scalar field $A\phi+(2z-2-\theta)B\zeta$ with
positive mass, driven by the source field $\zeta$.\ Since $\zeta$ is
also a positive mass scalar, small fluctuations do not contain any
unstable modes growing in time. Thus the general perturbation also is
stable. To elaborate a bit further, imagine long-wavelength modes of
$\phi, \zeta$ which are spatially uniform,
\ie\ $\phi=\phi(t),\ \zeta=\zeta(t)$. Now the linearized equations are
of the form\ $\ddot\phi+m^2_\phi\phi=0,\ \ddot\zeta+m^2_\zeta\zeta=0$,
so that these fields are effectively decoupled harmonic
oscillators. Then the general field is a driven oscillator, with the
driving force itself executing small oscillations: so there are no
unstable modes growing in time. It is important to note that the
positivity of the mass term coefficients and the stability they imply
stems from the energy conditions and asymptotic boundary conditions,
which force $z>1$ and $2z-2-\theta>0$ for generic $z,\theta$ values.\\
It is worth noting that for fixed $\zeta$, the relative sizes of the
dilaton $\phi$ and hvLif scalar $\psi$ perturbations are\
${\psi\over\phi} \sim\ {L^2\over r_0^2}\, ({r_0\over r_{hv}})^{2\theta}\ll
{L^2\over r_0^2}$\ \ for \ $\theta>0$\ since\ ${r_0\over r_{hv}}\ll 1$\,.}
\end{itemize}

It is worth comparing this analysis with that for the higher
dimensional theory discussed earlier in (\ref{VeffPsiAction}),
(\ref{VeffPsiStab}): the scalar $\Psi$ has a canonical kinetic term
and the equation governing small fluctuations of $\Psi$ about the
attractor point acquires a mass term from ${\del^2U\over\del\Psi^2}$,
whose positivity dictates the stability of the attractor point. For a
theory with two scalars $\phi_1,\phi_2$ with canonical kinetic terms,
the stability of the linearized fluctuations can again be studied by
studying the second derivative matrix of the potential
$U(\phi_1,\phi_2)$ or the Hessian $[{\del^2U\over\del\phi_i\del\phi_j}]$.
Positivity of the Hessian then translates to stability of the
attractor extremum.  In the present case however, the effective action
is (\ref{chbb2deffactionWeyl}), and the kinetic terms for $\Phi$,
$\Psi$ are not canonical: thus the naive Hessian analysis to study the
stability of $U(\Phi,\Psi)$ about the attractor point is not
valid. Instead we must analyze perturbations about the attractor
point, which are governed by the above equations. From these
equations, we see that the mass terms for the decoupled fields $\zeta$
and $\phi$ are positive.

In terms of $\phi$ and $\zeta$, the quadratic action becomes
\begin{equation}\label{chbb2dS2new}\begin{split}
	S_2=&\frac{1}{16\pi G_2}\int d^2x\,\frac{r_0^{2-2\theta}r_{hv}^{2\theta}}{L^2(2+z-\theta)(1+z-\theta)}\,\Big[8\phi\,\partial_+\partial_-\Omega+\frac{16}{(x^+-x^-)^2}\phi\Omega \\[3pt]
	& +\frac{r_0^{2-2\theta}r_{hv}^{2\theta}}{L^2(2+z-\theta)(1+z-\theta)}\Big((2z-2-\theta)\partial_+\zeta\partial_-\zeta-\frac{4(z-1)}{(x^+-x^-)^2}\zeta^2\Big) \\[3pt]
	& +\frac{L^2(2+z-\theta)(1+z-\theta)}{r_0^{2-2\theta}r_{hv}^{2\theta}}\Big(\frac{4(2z-2-\theta)}{(2-\theta)}\partial_+\phi\partial_-\phi-\frac{16(z+1-2\theta)}{(2-\theta)(x^+-x^-)^2}\phi^2\Big) \\[3pt]
	& +2\sqrt{\frac{2z-2-\theta}{2-\theta}}(\partial_+\zeta\partial_-\phi+\partial_-\zeta\partial_+\phi)-\frac{8(2z-2-\theta)}{\sqrt{2-\theta}}\frac{\zeta\phi}{(x^+-x^-)^2}\Big].
\end{split}\end{equation}
It can be checked that varying this action leads to the linearized equations
written in terms of $\phi,\zeta$ above.

\subsubsection{The Schwarzian}

In this section, we switch to Euclidean time $\tau=it$. From the linearized
equations \eqref{chbb2dlpsieqn},
we see that the dilaton fluctuation $\phi$ is decoupled from the
metric and scalar fluctuations $\Omega$ and $\psi$, as in the case of
the relativistic brane. So solving the equation for $\phi$ (\ie\ the Euclidean form of \eqref{chbb2dlpsieqn}) gives, as before,
\begin{equation}
	\phi=\frac{a+b\tau+c(\tau^2+\rho^2)}{\rho}\ ,
\end{equation}
where $a$, $b$, $c$ are independent constants. Substituting $\phi$ in the equation for $\psi$ in \eqref{chbb2dlpsieqn}, we can solve for the scalar perturbation
$\psi$. Using these solutions for $\phi$ and $\psi$ in the equation for $\Omega$ in \eqref{chbb2dlpsieqn}, we can solve for the metric perturbation
$\Omega$. We see that the $AdS_2$ metric gets corrected at the same
order as the dilaton and the scalar field.
The Euclidean on-shell (boundary) action obtained by using linearized field
equations in \eqref{chbb2dS2new} and changing to Euclidean time $\tau=it$ is
\begin{equation}\begin{split}\label{S2Esublead}
	S^E_2=&-\frac{1}{16\pi G_2}\int d\tau\sqrt{\gamma}\,n^{\mu}\frac{r_0^{2-2\theta}r_{hv}^{2\theta}}{L^2(2+z-\theta)(1+z-\theta)}\Big[4(\Omega\partial_{\mu}\phi-\phi\partial_{\mu}\Omega) \\
	& -\frac{(2z-2-\theta)}{\sqrt{2-\theta}}(\phi\partial_{\mu}\zeta+\zeta\partial_{\mu}\phi)-\frac{L^2(2+z-\theta)(1+z-\theta)}{r_0^{2-2\theta}r_{hv}^{2\theta}}\frac{2(2z-2-\theta)}{(2-\theta)}\phi\partial_{\mu}\phi \\
	& -\frac{r_0^{2-2\theta}r_{hv}^{2\theta}}{L^2(2+z-\theta)(1+z-\theta)}(2z-2-\theta)\zeta\partial_{\mu}\zeta\Big]\ .
\end{split}\end{equation}
The discussion of the Gibbons-Hawking term is very similar to that in
sec.~\ref{Schw:rel} so we will not be detailed.\
The Gibbons-Hawking boundary term for the Euclidean form of the bulk action
\eqref{chbb2deffactionWeyl} is
\begin{equation}
S_{GH}=-\frac{1}{8\pi G_2}\int d\tau\,\sqrt{\gamma}\,\Phi^2\,K\ ,
\end{equation}
arising as discussed in the case of the relativistic electric brane
earlier. As in sec.~\ref{Schw:rel}, we now redefine the dilaton
perturbation after rescaling the background value $\Phi_b$ out, so
that the perturbation satisfies\
${\Phi-\Phi_b\over\Phi_b} \equiv {\tilde\phi}\ll 1$.\ A similar
redefinition is appropriate for the hvLif scalar $\Psi$ as well\
(we have however retained the perturbations in
(\ref{chbb2dperturbations}) without this rescaling simply with a view
to not cluttering the resulting expressions).\
Then the perturbation, the background value
(\ref{chbb2dexads2}) and the entropy (\ref{entropyztheta}) are
\be\label{entropyztheta-2}
\Phi=\Phi_b\, (1+{\tilde\phi})\ ,\qquad\quad
\Phi_b^2=\Big(\frac{r_0}{r_{hv}}\Big)^{-\theta}\Big(\frac{r_0}{R}\Big)^2 ,
\qquad\quad S_{BH} = {\Phi_b^2\, V_2\over 4G_4} = {\Phi_b^2\over 4G_2}\ .
\ee
This gives
\begin{equation}\label{SchwEntrztheta}
  S_{GH}^{(1)}=-\frac{2\Phi_b^2}{8\pi G_2}\int d\tau\sqrt{\gamma}\
  {\tilde\phi}\, K\ \longrightarrow\
  -\frac{\Phi_b^2}{4\pi G_2}\int du\, \phi_r(u)\, \{\tau(u),u\}\ .
\end{equation}
In evaluating the last term, we take the boundary of $AdS_2$ as a
slightly deformed curve $(\tau(u),\rho(u))$ parametrized by the boundary
coordinate $u$, as discussed in \cite{Maldacena:2016upp} (reviewed in
\cite{Sarosi:2017ykf}),\ and expand the extrinsic curvature using the
outward unit normal $n^{\mu}$ to the boundary.
Expanding $S^{(1)}_{GH}$ leads to the action above, which contains the
Schwarzian derivative\
$Sch(\tau(u),u)=\{\tau(u),u\}={\tau'''\over \tau'}-{3\over 2} ({\tau''\over \tau'})^2$\,.\
The integral above pertains simply to the $AdS_2$ scale $L$, into which
the various length scales in the nonrelativistic theory have been absorbed.
We have also as before defined\ ${\tilde\phi}={\phi_r(u)\over\epsilon}$\
and\ $\sqrt{\gamma}={L\over\epsilon}$\,.

As for the relativistic brane sec.~\ref{Schw:rel} and
(\ref{SchwEntrz=1}), we note that the coefficient of the Schwarzian
effective action is proportional to the entropy
(\ref{entropyztheta}), (\ref{entropyztheta-2}) of the compactified
black brane, with $V_2$ finite. As in sec.~\ref{Schw:rel}, this
coefficient as the entropy arises after making the reasonable
definition of the dilaton perturbation as in (\ref{entropyztheta-2}),
scaling out the background $\Phi_b$.  The entropy now contains only
$\Phi_b$, which controls the transverse area. Since the entropy
captures the number of microstates of the unperturbed background, this
is akin to a central charge.

This is the leading term in the total boundary action $I_{bdy}=S^E_2+S_{GH}$.
The remaining terms in the expansion of $S_{GH}$ and $S^E_2$ are
all quadratic in perturbations and hence are subleading compared to
$S_{GH}^{(1)}$ which contains the dilaton perturbation alone at linear
order, as for the relativistic brane discussed earlier. This universal
behaviour is in accord with the general arguments in \eg\
\cite{Maldacena:2016upp}.

Thus overall, expanding in the perturbations $\tilde{\phi}$, $\Omega$,
$\psi$, we have $I=S^E+S_{GH}=I_0+I_1+I_2+\dots$, where
\begin{equation}
  I_0=-\frac{\Phi_b^2}{16\pi G_2}
  \left(\int d^2x \sqrt{g}\,\mathcal{R} +2\int_{{\tiny bndry}}
  \sqrt{\gamma}\,K\right)\ ,
\end{equation}
is the background action (see (\ref{S0ztheta})): here $\Psi_b$ is
constant and it can be checked that $U(\Phi_b,\Psi_b)=0$. This is a
topological term and gives the extremal entropy, very similar to the
detailed discussion for the relativistic brane sec.~\ref{Schw:rel}.
The linear terms are contained in
\begin{eqnarray}
  && I_1=-\frac{2\Phi_b^2}{16\pi G_2}\int d^2x\,\sqrt{g}\,
  \tilde{\phi}\,\Big(\mathcal{R}-\frac{\partial U}{\partial\Phi^2}-\frac{1}{2}(\partial\Psi_b)^2\Big)
  -\frac{2\Phi_b^2}{8\pi G_2}\int_{bndry} \sqrt{\gamma}\,\tilde{\phi}\,K \nonumber\\
  && \qquad\qquad -\frac{1}{16\pi G_2}\int d^2x\,\sqrt{g}\Big(-\frac{\Phi_b^2}{2}\partial_{\mu}\Psi_b\partial^{\mu}\psi-\psi\frac{\partial U}{\partial\Psi}\Big)\ .
\end{eqnarray}
On the $AdS_2$ background with a constant dilaton $\Phi_b$ and a
constant hvLif scalar field $\Psi_b$, we get $\frac{\partial
  U}{\partial\Phi^2}|_{(\Phi_b,\Psi_b)}=-\frac{2}{L^2}$ and the second
line in the expression for $I_1$ above vanishes by the $\Psi$ equation
in \eqref{chbb2dequiveoms}. Thus, $I_1$ reduces to
\begin{equation}
	I_1=-\frac{2\Phi_b^2}{16\pi G_2}\int d^2x\,\sqrt{g}\,
	\tilde{\phi}\,\Big(\mathcal{R}+\frac{2}{L^2}\Big)-\frac{2\Phi_b^2}{8\pi G_2}\int_{bndry}\sqrt{\gamma}\,\tilde{\phi}\,K\ ,
\end{equation}
which is the Jackiw-Teitelboim theory. The fluctuations of the scalar
$\Psi$ now propagate on the fixed $AdS_2$ background at this
order. However we see as in sec.~\ref{Schw:rel} that there are various
subleading terms at quadratic order ((\ref{S2Esublead}) and from the
Gibbons-Hawking term, see (\ref{I2z=1}), as well as possible
counterterms), containing the perturbations to the dilaton $\Phi$,
metric and scalar $\Psi$, which all mix (at the same order as the
metric): the fluctuation spectrum is stable for physically sensible
theories satisfying the energy conditions as we have seen.  These
encode information about the regularization of the $AdS_2$ theory by
the particular higher dimensional hvLif theory.

\subsubsection{More general perturbations}\label{sec:moregenpert}

In the above analysis we have restricted ourselves to the dimensional
reduction of perturbations to only those components of fields (metric,
gauge fields, scalar) which have non-trivial background values in the
higher dimensional theory. More generally, considering the dimensional
reduction of perturbations to all the components of all the fields
(some of which have trivial background values) gives
\be
h_{MN}\ \ra\ h_{\mu\nu} ,\ \ \ h_{\mu i} ,\ \ \ h_{ij}\ ;\qquad\ \ 
A_M^{(1,2)}\ \ra\ A^{(1,2)}_\mu ,\ \ \ A_i^{(1,2)}\ ;\qquad\ \ 
\phi\ra \phi\ ,
\ee
\ie\ tensor, vector and scalar perturbations in the 2-dimensional theory\
(note that the 2-dim dilaton is $g_{xx}$).
For instance this includes the shear perturbation $h_{xy}$ in the higher
dimensional theory as well the spatial components of the gauge fields 
$A_{1\,i}$, $A_{2\,i}$ for $i=x,y$ which reduce respectively to a non-minimally
coupled scalar ($h=g^{(4)\,xx}h_{xy}$) and minimally coupled scalars
$A_{1\,i}=\chi^{(1)}_i$, $A_{2\,i}=\chi^{(2)}_i$ in the
$2$-dimensional theory. 
The terms in the full $2$-dimensional action which govern these
perturbations are
\begin{equation}
	S=\frac{1}{16\pi G}\int d^2x \sqrt{-g}\Big[\dots -\frac{\Phi^2}{2}(\partial h)^2-\frac{e^{\lambda_1\Psi}}{2}(\partial\chi^{(1)}_i)^2-\frac{e^{\lambda_2\Psi}}{2}(\partial\chi^{(2)}_i)^2 \Big]\ .
\end{equation}
The terms involving $h_{xy}$ arise from the higher dimensional Ricci
scalar and so contain the overall dilaton factor $\Phi^2$ under reduction
to 2-dimensions. The linearized equations for $h_{xy}$ in the higher
dimensional theory in \eg\ \cite{Kolekar:2016pnr} can be dimensionally
reduced to 2-dimensions: at zero momentum, this is consistent with the
Kaluza-Klein ansatz for reduction and the action above.
Expanding these terms around the background $AdS_2$, the leading
contributions from these terms appear at quadratic order in perturbations
\begin{equation}
	S_2=\frac{1}{16\pi G}\int d^2x \sqrt{-g}\Big[\dots -\frac{\Phi_b^2}{2}(\partial h)^2-\frac{e^{\lambda_1\Psi_b}}{2}(\partial\chi^{(1)}_i)^2-\frac{e^{\lambda_2\Psi_b}}{2}(\partial\chi^{(2)}_i)^2 \Big]\ .
\end{equation}
These are subleading compared to $S_{GH}^{(1)}$ and thus do not
contribute to the Schwarzian.

\section{On a null reduction of the charged $AdS_5$ black brane}

In \cite{Narayan:2012hk} (see also \cite{Singh:2012un}), it was argued
that the null reduction of $AdS$ plane waves, highly boosted limits of
uncharged black branes, gives rise to hvLif theories with certain
specific $z,\theta$ exponents.  The lower dimensional hvLif gauge
field and scalar arise as the KK gauge field and scalar under
$x^+$-reduction. One might imagine that considering such a null
reduction of the charged relativistic black brane might be interesting
along these lines.  In this section, we describe an attempt to obtain
the charged hvLif black branes here by a null $x^+$-reduction of the
charged relativistic black brane in one higher dimension.
Unfortunately this turns out to be close, but not quite on the nose:
while the charge electric gauge field upstairs does give rise to an
electric field in the lower dimensional theory, it also leads to an
additional background scalar profile. It would be interesting to
understand if this can be refined further.

The action for a charged $AdS_5$ black brane \cite{Hartnoll:2016apf}
is\footnote{In this section, $r\rightarrow 0$ is the boundary.}
\begin{equation}\label{cads5bbaction}
	S=\frac{1}{2\kappa^2}\int d^5x\sqrt{-g}\Big[\mathcal{R}-2\Lambda-\frac{2\kappa^2}{e^2}\frac{F^2}{4}\Big]\ .
\end{equation}
The charged $AdS_5$ black brane metric is
\begin{equation}\label{cads5bbmetric}
	ds^2=\frac{L^2}{r^2}\Big(-f(r)dt^2+\frac{dr^2}{f(r)}+dx_1^2+dx_2^2+dx_3^2 \Big)\ ,
\end{equation}
\begin{equation} 
f(r)=1-\Big(1+\frac{r_0^2\mu^2}{\gamma^2}\Big(1-\frac{r^2}{r_0^2}\Big)\Big)\Big(\frac{r}{r_0}\Big)^4\ , \qquad\quad \gamma^2=\frac{3e^2L^2}{2\kappa^2}\ ,
\end{equation}
where the horizon is at $r=r_0$ and the boundary at $r\rightarrow
0$. The gauge field $A_t$, charge density $\rho$ and temperature are
\begin{equation}
	A_t=\mu\left(1-\Big(\frac{r}{r_0}\Big)^2\right)\ , \qquad \rho=\frac{2L}{e^2r_0^2}\mu\ ,\qquad 	T=\frac{1}{4\pi r_0}\left(4-2\frac{r_0^2\mu^2}{\gamma^2}\right)\ .
\end{equation}
Transforming to lightcone coordinates, $x^{\pm}=\frac{t\pm
  x_3}{\sqrt{2}}$ and performing a boost $x^{\pm}\rightarrow
\lambda^{\pm}x^{\pm}$, the metric becomes
\begin{equation}\label{cads5bbmetricboosted}
	ds^2=\frac{L^2}{r^2}\left(-f(r)\left(\frac{\lambda dx^+ +\lambda^{-1}dx^-}{\sqrt{2}}\right)^2+\left(\frac{\lambda dx^+ -\lambda^{-1}dx^-}{\sqrt{2}}\right)^2+\frac{dr^2}{f(r)}+dx_1^2+dx_2^2\right)\ .
\end{equation}
Completing squares in $dx^+$, we get
\begin{equation}\label{4dmetric}\begin{split}
ds^2&=-\frac{2L^2r_0^2f(r)}{\lambda^2r^6\Big(1+\frac{r_0^2\mu^2}{\gamma^2}\Big(1-\frac{r^2}{r_0^2}\Big)\Big)}(dx^-)^2+\frac{L^2}{r^2}\left(\frac{dr^2}{f(r)}+dx_1^2+dx_2^2\right) \\
&\qquad\quad +\frac{L^2\lambda^2r^2}{2r_0^4}\Big(1+\frac{r_0^2\mu^2}{\gamma^2}\Big(1-\frac{r^2}{r_0^2}\Big)\Big)(dx^+ +\mathcal{A}_-dx^-)^2\ ,
\end{split}\end{equation}
\begin{equation}\label{hvLgaugefield}
	\mathcal{A}_-=\frac{-1+\frac{r^4}{2r_0^4}\Big(1+\frac{r_0^2\mu^2}{\gamma^2}\Big(1-\frac{r^2}{r_0^2}\Big)\Big)}{\frac{\lambda^2r^4}{2r_0^4}\Big(1+\frac{r_0^2\mu^2}{\gamma^2}\Big(1-\frac{r^2}{r_0^2}\Big)\Big)}\ .
\end{equation}
The first line in \eqref{4dmetric} after incorporating the conformal factor
from $x^+$-reduction leads approximately to the 4-dim hvLif metric with
$z=3$, $\theta=1$, in the vicinity of $r\rightarrow 0$ and $r\rightarrow r_0$.
The KK-gauge field becomes the $F_1$ gauge field in the lower dimensional
theory: its form becomes that of $F_{1rt}$ only in the vicinity of the
horizon $r\ra r_0$, giving\
$A_{1-}\equiv A_{1t}\sim -{1\over (\lambda^2/r_0^4) r^4} + {1\over\lambda^2}$\,,
where we hold ${\lambda^2\over r_0^4}$ fixed which preserves the first
term (while the 2nd term dies).
This reduction to hvLif is exact if $\mu=0$, as in \cite{Narayan:2012hk}
for zero temperature (and \cite{Singh:2012un,Kolekar:2016pnr} for
finite temperature).

Likewise the $A_t\equiv A_{2t}$ gauge field giving charge becomes
in the lower dimensional theory
\be
A_{2+} = \lambda A_{2t}\ ,\qquad
A_{2-} = {1\over\lambda} A_{2t} \ra A_{2t}^{4d}\ .
\ee
Scaling the chemical potential as $\mu\ra {\mu\over\lambda}=fixed$, we
obtain precisely the gauge field profile for $A_{2t}$: however $A_{2+}$
survives as a scalar background in the lower dimensional theory.

It can also be seen that the relativistic brane action
(\ref{cbb4daction}) gives rise upon $x^+$-reduction to the hvLif
action (\ref{chbb4daction}), upto the extra scalar arising from
$A_{2+}$. It would be interesting look for refinements of the
discussion here, towards decoupling this extra scalar.

\section{Discussion}

We have studied dilaton-gravity theories in 2-dimensions obtained by
dimensional reduction of certain families of extremal charged
hyperscaling violating Lifshitz black branes in
Einstein-Maxwell-scalar theories with an extra gauge field in
4-dimensions. We have argued that the near horizon $AdS_2$ backgrounds
here can be obtained in equivalent theories of 2-dim dilaton-gravity
with an extra scalar, descending from the higher dimensional scalar,
and an interaction potential with the dilaton. A simple subcase is the
relativistic black brane with $z=1, \theta=0$\ (which has no extra
scalar), which we have analysed in detail. Studying linearized fluctuations
of the metric, dilaton and the extra scalar about these $AdS_2$
backgrounds suggests stability of the attractor background
generically. This is correlated with the requirements imposed by
the energy conditions on these backgrounds.
From the study of small fluctuations, we have seen that the leading
corrections to $AdS_2$ arise at linear order in the dilaton
perturbation resulting in a Schwarzian derivative effective action
from the Gibbons-Hawking term, and Jackiw-Teitelboim theory at leading
order.  We have also seen that the coefficient of the Schwarzian
derivative term, (\ref{SchwEntrz=1}), (\ref{SchwEntrztheta}), is
proportional to the entropy of the (compactified) extremal black
branes after defining the perturbations by scaling out the background
values (\ref{dilpertz=1}), (\ref{entropyztheta-2}): this being the
number of microstates of the unperturbed background is thus akin to a
central charge. The background entropy arises automatically as a
topological term from the compactification. There are of course
various subleading terms in the action at quadratic order which mix at
the same order as the metric: these encode information on the higher
dimensional realization of these $AdS_2$ backgrounds.

We have explored certain classes of such extremal backgrounds: it
would be interesting to understand the space of such $AdS_2$ theories
in a more systematic manner. One might imagine that the parameters in
these theories, for instance the dynamical exponents, are reflected in
the spectrum of correlation functions, thus distinguishing the
specific ultraviolet regularization of the $AdS_2$ regimes. This
requires better understanding of the subleading terms beyond the
Schwarzian, which in turn requires a systematic treatment of
counterterms and holographic renormalization. We hope to explore these
further.

From the point of view of the dual theories, it would seem that the
present 2-dim backgrounds are dual to 1-dimensional theories arising
from $T^2$ compactifications of the dual field theories. It would be
interesting to understand these better, in part towards possibly
exploring parallels with the SYK models \cite{Sachdev:1992fk,
  Kitaev-talks:2015}, discussed more recently in \eg\
\cite{Polchinski:2016xgd,Maldacena:2016upp,Maldacena:2016hyu,Kitaev:2017awl}
and related SYK/tensor models\ (see \eg\ \cite{Fu:2016yrv}).

\vspace{7mm}

{\footnotesize \noindent {\bf Acknowledgements:}\ \ It is a pleasure
  to thank Sumit Das, Vladimir Rosenhaus and Amitabh Virmani for
  comments on a draft.  KK also thanks Alok Laddha, Shiraz Minwalla
  and Vladimir Rosenhaus for discussions. The research of KK was
  supported in part by the International Centre for Theoretical
  Sciences (ICTS), Bangalore, during a visit for participating in the
  program - Kavli Asian Winter School (KAWS) on Strings, Particles and
  Cosmology 2018 (Code: ICTS/Prog-KAWS2018/01). KK also thanks the
  organizers of the National Strings Meeting 2017 (NISER Bhubaneshwar)
  and KN thanks the organizers of the IV Saha Theory Workshop on
  Modern Aspects of String Theory, Kolkata, 2018, for hospitality
  while this work was in progress.  This work is partially supported
  by a grant to CMI from the Infosys Foundation. }

\appendix
\section{Some details}

\noindent {\bf Relativistic electric black brane:}\ \ The Einstein
equation and the dilaton equation from the action \eqref{cbb2dFaction}
are
\begin{eqnarray}\label{cbb2deoms}
&&  g_{\mu\nu}\nabla^2\Phi^2-\nabla_{\mu}\nabla_{\nu}\Phi^2+\frac{g_{\mu\nu}}{2}\Big(2\Lambda\Phi+\frac{16\pi G_2V_2\Phi^3F_{\mu\nu}F^{\mu\nu}}{4}\Big)-\frac{16\pi G_2V_2\Phi^3}{2}F_{\mu\rho}F_{\nu}^{\ \rho} = 0\ , \nonumber\\
&&  \mathcal{R}-\frac{\Lambda}{\Phi}-(6\pi G_2)V_2\Phi F_{\mu\nu}F^{\mu\nu}
  = 0\ .
\end{eqnarray}

\noindent \underline{{\bf Charged hvLif black brane}}\\
{\bf Effective scalar potential in $4$-dimensional hvLif black brane and
its derivatives:}\ \ 
The first and second derivatives of the effective scalar potential in
$4$-dimensional charged hvLif black brane are
\bea\label{chbb4dVeff'}
  && \frac{\partial V_{eff}}{\partial\Psi}=-\frac{\gamma(2+z-\theta)(1+z-\theta)e^{\gamma(\Psi-\Psi_0)}}{R^{2-2\theta}r_{hv}^{2\theta}} \\
  && \quad\qquad\  \ -\frac{1}{g_{xx}^2}\Big(\frac{\lambda_1(z-1)(2+z-\theta)r_{hv}^{2\theta-4}R^{2-2\theta}}{e^{\lambda_1(\Psi-\Psi_0)}}+\frac{\lambda_2(2-\theta)(z-\theta)Q^2r_{hv}^{2z-2}R^{-4z-2+2\theta}}{e^{\lambda_2(\Psi-\Psi_0)}}\Big), \nonumber
\eea
\bea\label{chbb4dVeff''}
  && \frac{\partial^2 V_{eff}}{\partial\Psi^2}=-\frac{\gamma^2(2+z-\theta)(1+z-\theta)e^{\gamma(\Psi-\Psi_0)}}{R^{2-2\theta}r_{hv}^{2\theta}} \\
  && \quad\qquad\ \ +\frac{1}{g_{xx}^2}\Big(\frac{\lambda_1^2(z-1)(2+z-\theta)r_{hv}^{2\theta-4}R^{2-2\theta}}{e^{\lambda_1(\Psi-\Psi_0)}}+\frac{\lambda_2^2(2-\theta)(z-\theta)Q^2r_{hv}^{2z-2}R^{-4z-2+2\theta}}{e^{\lambda_2(\Psi-\Psi_0)}}\Big). \nonumber
\eea
Differentiating $V_{eff}$ $n$ times, we get
\bea
	&& \frac{\partial^n V_{eff}}{\partial\Psi^n}=-\frac{\gamma^n(2+z-\theta)(1+z-\theta)e^{\gamma(\Psi-\Psi_0)}}{R^{2-2\theta}r_{hv}^{2\theta}} \\
	&& \quad\qquad\ \ +\frac{(-)^n}{g_{xx}^2}\Big(\frac{\lambda_1^{n}(z-1)(2+z-\theta)r_{hv}^{2\theta-4}R^{2-2\theta}}{e^{\lambda_1(\Psi-\Psi_0)}}+\frac{\lambda_2^{n}(2-\theta)(z-\theta)Q^2r_{hv}^{2z-2}R^{-4z-2+2\theta}}{e^{\lambda_2(\Psi-\Psi_0)}}\Big),\nonumber
\eea
which at the extremal point becomes
\begin{equation}\label{chbb4dVeff'n}
  \frac{\partial^nV_{eff}}{\partial\Psi^n}=\frac{r_0^{\theta}(2+z-\theta)}{r_{hv}^{\theta}R^2}\Big[\frac{-\theta^n(1+z-\theta)+(-)^n(\theta-4)^n(z-1)}{(2-\theta)^{\frac{n}{2}}(2z-2-\theta)^{\frac{n}{2}}}+\frac{(-)^n(2z-2-\theta)^{\frac{n}{2}}(2-\theta)}{(2-\theta)^{\frac{n}{2}}}\Big].
\end{equation}
At $z=1$, $\theta\neq 0$, we see that $\frac{\partial^n V_{eff}}{\partial\Psi^n}=0$ $\forall\ n$ at the extremal point.

\vspace{1mm}

\noindent {\bf Dimensional reduction to $2$-dimensions:}\ \ 
The 2-dim action obtained by reducing \eqref{chbb4daction} on $T^2$ is
(retaining only fields with background profiles)
\begin{equation}\label{chbb2daction}\begin{split}
  S=\int d^2x\sqrt{-g^{(2)}}&\Big[\frac{1}{16\pi G_2}\Big(\Phi^2\mathcal{R}^{(2)}+2\partial_{\mu}\Phi\partial^\mu\Phi-\frac{\Phi^2}{2}\partial_{\mu}\Psi\partial^{\mu}\Psi+V\Phi^2 \\
  & -\frac{\Phi^2}{4}Z_1 F_{1\,\mu\nu}F_1^{\mu\nu}\Big)-\frac{V_2\Phi^2}{4}Z_2 F_{2\,\mu\nu}F_2^{\mu\nu}\Big] \ ,
\end{split}\end{equation}

\noindent {\bf Equations of motion from $2$-dimensional action
  \eqref{chbb2dactionWeyl}:}\ \ 
The equations of motion obtained by varying the action
\eqref{chbb2dactionWeyl} are
\begin{eqnarray}\label{chbb2dnoWeyleoms}
&&  g_{\mu\nu}\nabla^2\Phi^2-\nabla_{\mu}\nabla_{\nu}\Phi^2+\frac{g_{\mu\nu}}{2}\Big(\frac{\Phi^2}{2}(\partial\Psi)^2-V\Phi+\frac{\Phi^3}{4}(Z_1 (F_1)^2+16\pi G_2V_2Z_2 (F_2)^2)\Big) \nonumber\\
&&  \hspace{20mm} -\frac{\Phi^2}{2}\partial_{\mu}\Psi\partial_{\nu}\Psi-\frac{\Phi^3}{2}(Z_1 F_{1\,\mu\rho}F_{1\nu}^{\ \ \rho}+16\pi G_2V_2Z_2 F_{2\,\mu\rho}F_{2\nu}^{\ \ \rho}) = 0\ , \nonumber\\
&& \mathcal{R}-\frac{1}{2}(\partial\Psi)^2+\frac{V}{2\Phi}-\frac{3}{8}\Phi(Z_1 (F_1)^2+16\pi G_2V_2Z_2 (F_2)^2) = 0\ , \nonumber\\
&&  \frac{1}{\sqrt{-g}}\partial_{\mu}(\sqrt{-g}\Phi^2\partial^{\mu}\Psi)+\gamma V\Phi-\frac{\Phi^3}{4}(\lambda_1 Z_1(F_1)^2+\lambda_2 16\pi G_2V_2Z_2(F_2)^2)
  = 0\ .
\end{eqnarray}
The equations of motion \eqref{chbb2dequiveoms} in conformal gauge and
in lightcone coordinates are
\begin{eqnarray}\label{chbb2deomsclc}
-e^{2\omega}\partial_{\pm}(e^{-2\omega}\partial_{\pm}\Phi^2)-\frac{\Phi^2}{2}\partial_{\pm}\Psi\partial_{\pm}\Psi &=& 0\ , \nonumber\\
\partial_+\partial_-\Phi^2-\frac{e^{2\omega}}{4}U &=& 0\ ,
  \nonumber\\
4\partial_+\partial_-\omega+\partial_+\Psi\partial_-\Psi-\frac{ e^{2\omega}}{2}\frac{\partial U}{\partial(\Phi^2)} &=& 0\ , \nonumber\\
\partial_+(\Phi^2\partial_-\Psi)+\partial_-(\Phi^2\partial_+\Psi)+\frac{e^{2\omega}}{2}\frac{\partial U}{\partial\Psi} &=& 0\ .\ \ \ 
\end{eqnarray}
Expanding the constraint equations in the first line of
\eqref{chbb2deomsclc} to linear order in perturbations
\eqref{chbb2dperturbations} gives
\begin{equation}\label{chbb2deomsclcconstraint1}
	\partial_{\pm}\partial_{\pm}\phi \pm \frac{2}{(x^+ -x^-)}\partial_{\pm}\phi=0\ ,
\end{equation}
the other terms vanishing at linear order.
To see that these linearized constraint equations are consistent with the
linearized equations \eqref{chbb2dlpsieqn}, we differentiate the $++$
constraint equation with respect to $x^-$ to get
\begin{equation}\label{chbb2deomsclcconstraint2}
	\partial_+(\partial_+\partial_-\phi)+\frac{2}{(x^+ -x^-)}\partial_+\partial_-\phi+\frac{2}{(x^+ -x^-)^2}\partial_+\phi=0\ ,
\end{equation}
which is satisfied after using the equation for $\phi$ in \eqref{chbb2dlpsieqn}. Similarly differentiating the $--$ constraint equation with respect to $x^+$, we can show that the resulting equation is satisfied upon substituting the equation for $\phi$ in \eqref{chbb2dlpsieqn}.


\begin{thebibliography}{}  

\footnotesize{
    
\bibitem{Almheiri:2014cka} 
A.~Almheiri and J.~Polchinski,
``Models of AdS$_{2}$ backreaction and holography,''
JHEP {\bf 1511}, 014 (2015)
doi:10.1007/JHEP11(2015)014
[arXiv:1402.6334 [hep-th]].
  
  \bibitem{Maldacena:2016upp} 
  J.~Maldacena, D.~Stanford and Z.~Yang,
  ``Conformal symmetry and its breaking in two dimensional Nearly Anti-de-Sitter space,''
  PTEP {\bf 2016}, no. 12, 12C104 (2016)
  doi:10.1093/ptep/ptw124
  [arXiv:1606.01857 [hep-th]].
	
\bibitem{Jensen:2016pah} 
  K.~Jensen,
  ``Chaos in AdS$_2$ Holography,''
  Phys.\ Rev.\ Lett.\  {\bf 117}, no. 11, 111601 (2016)
  doi:10.1103/PhysRevLett.117.111601
  [arXiv:1605.06098 [hep-th]].

\bibitem{Engelsoy:2016xyb} 
J.~Engelsoy, T.~G.~Mertens and H.~Verlinde,
``An investigation of AdS$_{2}$ backreaction and holography,''
JHEP {\bf 1607}, 139 (2016)
doi:10.1007/JHEP07(2016)139
[arXiv:1606.03438 [hep-th]].

\bibitem{Almheiri:2016fws} 
A.~Almheiri and B.~Kang,
``Conformal Symmetry Breaking and Thermodynamics of Near-Extremal Black Holes,''
	JHEP {\bf 1610}, 052 (2016)
	doi:10.1007/JHEP10(2016)052
	[arXiv:1606.04108 [hep-th]].

\bibitem{Maldacena:1998uz} 
  J.~M.~Maldacena, J.~Michelson and A.~Strominger,
  ``Anti-de Sitter fragmentation,''
  JHEP {\bf 9902}, 011 (1999)
  doi:10.1088/1126-6708/1999/02/011
  [hep-th/9812073].

\bibitem{Sen:2008vm}
  A.~Sen,
  ``Quantum Entropy Function from AdS(2)/CFT(1) Correspondence,''
  Int.\ J.\ Mod.\ Phys.\ A {\bf 24}, 4225 (2009)
  doi:10.1142/S0217751X09045893
  [arXiv:0809.3304 [hep-th]].

  \bibitem{Sachdev:1992fk} 
  S.~Sachdev and J.~Ye,
  ``Gapless spin fluid ground state in a random, quantum Heisenberg magnet,''
  Phys.\ Rev.\ Lett.\  {\bf 70}, 3339 (1993)
  doi:10.1103/PhysRevLett.70.3339
  [cond-mat/9212030].

\bibitem{Kitaev-talks:2015}
  A.~Kitaev, \emph{``A simple model of quantum holography''},
http://online.kitp.ucsb.edu/online/entangled15/kitaev/,\ \
http://online.kitp.ucsb.edu/online/entangled15/kitaev2/,
Talks at the KITP, Santa Barbara, 2015.

  \bibitem{Kitaev:2017awl} 
  A.~Kitaev and S.~J.~Suh,
  ``The soft mode in the Sachdev-Ye-Kitaev model and its gravity dual,''
  arXiv:1711.08467 [hep-th].

  \bibitem{Sarosi:2017ykf} 
  G.~Sarosi,
  ``AdS2 holography and the SYK model,''
  arXiv:1711.08482 [hep-th].

\bibitem{Ferrara:1995ih} 
S.~Ferrara, R.~Kallosh and A.~Strominger,
``N=2 extremal black holes,''
Phys.\ Rev.\ D {\bf 52}, R5412 (1995)
doi:10.1103/PhysRevD.52.R5412
[hep-th/9508072].

  \bibitem{Goldstein:2005hq} 
  K.~Goldstein, N.~Iizuka, R.~P.~Jena and S.~P.~Trivedi,
  ``Non-supersymmetric attractors,''
  Phys.\ Rev.\ D {\bf 72}, 124021 (2005)
  doi:10.1103/PhysRevD.72.124021
  [hep-th/0507096].
  
  \bibitem{Tripathy:2005qp} 
  P.~K.~Tripathy and S.~P.~Trivedi,
  ``Non-supersymmetric attractors in string theory,''
  JHEP {\bf 0603}, 022 (2006)
  doi:10.1088/1126-6708/2006/03/022
  [hep-th/0511117].

  \bibitem{Hartnoll:2016apf} 
  S.~A.~Hartnoll, A.~Lucas and S.~Sachdev,
  ``Holographic quantum matter,''
  arXiv:1612.07324 [hep-th].

  \bibitem{Tarrio:2011de}
  J.~Tarrio and S.~Vandoren,
  ``Black holes and black branes in Lifshitz spacetimes,''
  JHEP {\bf 1109}, 017 (2011)
  doi:10.1007/JHEP09(2011)017
  [arXiv:1105.6335 [hep-th]].

  \bibitem{Alishahiha:2012qu} 
  M.~Alishahiha, E.~O Colgain and H.~Yavartanoo,
  ``Charged Black Branes with Hyperscaling Violating Factor,''
  JHEP {\bf 1211}, 137 (2012)
  doi:10.1007/JHEP11(2012)137
  [arXiv:1209.3946 [hep-th]].

\bibitem{Bueno:2012sd} 
  P.~Bueno, W.~Chemissany, P.~Meessen, T.~Ortin and C.~S.~Shahbazi,
  JHEP {\bf 1301}, 189 (2013)
  doi:10.1007/JHEP01(2013)189
  [arXiv:1209.4047 [hep-th]].

\bibitem{Cvetic:2016eiv} 
  M.~Cveti\v{c} and I.~Papadimitriou,
  ``AdS$_{2}$ holographic dictionary,''
  JHEP {\bf 1612}, 008 (2016)
  Erratum: [JHEP {\bf 1701}, 120 (2017)]
  doi:10.1007/JHEP12(2016)008, 10.1007/JHEP01(2017)120
  [arXiv:1608.07018 [hep-th]].

\bibitem{Castro:2008ms} 
  A.~Castro, D.~Grumiller, F.~Larsen and R.~McNees,
  ``Holographic Description of AdS(2) Black Holes,''
  JHEP {\bf 0811}, 052 (2008)
  doi:10.1088/1126-6708/2008/11/052
  [arXiv:0809.4264 [hep-th]].
  
\bibitem{Castro:2014ima} 
  A.~Castro and W.~Song,
  ``Comments on AdS$_2$ Gravity,''
  arXiv:1411.1948 [hep-th].
  
\bibitem{Das:2017pif} 
  S.~R.~Das, A.~Jevicki and K.~Suzuki,
  ``Three Dimensional View of the SYK/AdS Duality,''
  JHEP {\bf 1709}, 017 (2017)
  doi:10.1007/JHEP09(2017)017
  [arXiv:1704.07208 [hep-th]].
  
\bibitem{Taylor:2017dly} 
  M.~Taylor,
  ``Generalized conformal structure, dilaton gravity and SYK,''
  JHEP {\bf 1801}, 010 (2018)
  doi:10.1007/JHEP01(2018)010
  [arXiv:1706.07812 [hep-th]].

  \bibitem{Gaikwad:2018dfc} 
  A.~Gaikwad, L.~K.~Joshi, G.~Mandal and S.~R.~Wadia,
  ``Holographic dual to charged SYK from 3D Gravity and Chern-Simons,''
  arXiv:1802.07746 [hep-th].

  \bibitem{Nayak:2018qej} 
  P.~Nayak, A.~Shukla, R.~M.~Soni, S.~P.~Trivedi and V.~Vishal,
  ``On the Dynamics of Near-Extremal Black Holes,''
  arXiv:1802.09547 [hep-th].

\bibitem{Cadoni:2017dma} 
  M.~Cadoni, M.~Ciulu and M.~Tuveri,
  ``Symmetries, Holography and Quantum Phase Transition in Two-dimensional Dilaton AdS Gravity,''
  arXiv:1711.02459 [hep-th].

\bibitem{Jackiw:1984je} 
R.~Jackiw,
``Lower Dimensional Gravity,''
Nucl.\ Phys.\ B {\bf 252}, 343 (1985).
doi:10.1016/0550-3213(85)90448-1

\bibitem{Teitelboim:1983ux}
C.~Teitelboim,
``Gravitation and Hamiltonian Structure in Two Space-Time Dimensions,''
Phys.\ Lett.\  {\bf 126B}, 41 (1983).
doi:10.1016/0370-2693(83)90012-6

  \bibitem{Narayan:2012hk} 
  K.~Narayan,
  ``On Lifshitz scaling and hyperscaling violation in string theory,''
  Phys.\ Rev.\ D {\bf 85}, 106006 (2012)
  doi:10.1103/PhysRevD.85.106006
  [arXiv:1202.5935 [hep-th]].

  \bibitem{DHoker:2009mmn} 
  E.~D'Hoker and P.~Kraus,
  ``Magnetic Brane Solutions in AdS,''
  JHEP {\bf 0910}, 088 (2009)
  doi:10.1088/1126-6708/2009/10/088
  [arXiv:0908.3875 [hep-th]].

  \bibitem{Kolekar:2016pnr} 
  K.~S.~Kolekar, D.~Mukherjee and K.~Narayan,
  ``Hyperscaling violation and the shear diffusion constant,''
  Phys.\ Lett.\ B {\bf 760}, 86 (2016)
  [arXiv:1604.05092 [hep-th]],\
  ``Notes on hyperscaling violating Lifshitz and shear diffusion,''
  Phys.\ Rev.\ D {\bf 96}, no. 2, 026003 (2017)
  [arXiv:1612.05950 [hep-th]];\
  D.~Mukherjee and K.~Narayan,
  ``Hyperscaling violation, quasinormal modes and shear diffusion,''
  JHEP {\bf 1712}, 023 (2017)
  [arXiv:1707.07490 [hep-th]].

\bibitem{Singh:2012un} 
H.~Singh,
``Lifshitz/Schr\'odinger Dp-branes and dynamical exponents,''
JHEP {\bf 1207}, 082 (2012)
doi:10.1007/JHEP07(2012)082
[arXiv:1202.6533 [hep-th]].

  \bibitem{Polchinski:2016xgd} 
  J.~Polchinski and V.~Rosenhaus,
  ``The Spectrum in the Sachdev-Ye-Kitaev Model,''
  JHEP {\bf 1604}, 001 (2016)
  doi:10.1007/JHEP04(2016)001
  [arXiv:1601.06768 [hep-th]].

  \bibitem{Maldacena:2016hyu} 
  J.~Maldacena and D.~Stanford,
  ``Remarks on the Sachdev-Ye-Kitaev model,''
  Phys.\ Rev.\ D {\bf 94}, no. 10, 106002 (2016)
  doi:10.1103/PhysRevD.94.106002
  [arXiv:1604.07818 [hep-th]].
  
\bibitem{Fu:2016yrv} 
  W.~Fu and S.~Sachdev,
  ``Numerical study of fermion and boson models with infinite-range random interactions,''
  Phys.\ Rev.\ B {\bf 94}, no. 3, 035135 (2016)
  doi:10.1103/PhysRevB.94.035135
  [arXiv:1603.05246 [cond-mat.str-el]];\
  A.~Jevicki, K.~Suzuki and J.~Yoon,
  ``Bi-Local Holography in the SYK Model,''
  JHEP {\bf 1607}, 007 (2016)
  doi:10.1007/JHEP07(2016)007
  [arXiv:1603.06246 [hep-th]];\
  Y.~Gu, X.~L.~Qi and D.~Stanford,
  ``Local criticality, diffusion and chaos in generalized Sachdev-Ye-Kitaev models,''
  JHEP {\bf 1705}, 125 (2017)
  doi:10.1007/JHEP05(2017)125
  [arXiv:1609.07832 [hep-th]];\
  D.~J.~Gross and V.~Rosenhaus,
  ``A Generalization of Sachdev-Ye-Kitaev,''
  JHEP {\bf 1702}, 093 (2017)
  doi:10.1007/JHEP02(2017)093
  [arXiv:1610.01569 [hep-th]];\
  M.~Berkooz, P.~Narayan, M.~Rozali and J.~Simón,
  ``Higher Dimensional Generalizations of the SYK Model,''
  JHEP {\bf 1701}, 138 (2017)
  doi:10.1007/JHEP01(2017)138
  [arXiv:1610.02422 [hep-th]];\
  E.~Witten,
  ``An SYK-Like Model Without Disorder,''
  arXiv:1610.09758 [hep-th];\
  R.~Gurau,
  ``The complete $1/N$ expansion of a SYK–like tensor model,''
  Nucl.\ Phys.\ B {\bf 916}, 386 (2017)
  doi:10.1016/j.nuclphysb.2017.01.015
  [arXiv:1611.04032 [hep-th]];\
  I.~R.~Klebanov and G.~Tarnopolsky,
  ``Uncolored random tensors, melon diagrams, and the Sachdev-Ye-Kitaev models,''
  Phys.\ Rev.\ D {\bf 95}, no. 4, 046004 (2017)
  doi:10.1103/PhysRevD.95.046004
  [arXiv:1611.08915 [hep-th]];\
  T.~Nishinaka and S.~Terashima,
  Nucl.\ Phys.\ B {\bf 926}, 321 (2018)
  doi:10.1016/j.nuclphysb.2017.11.012
  [arXiv:1611.10290 [hep-th]];\
  C.~Krishnan, S.~Sanyal and P.~N.~Bala Subramanian,
  ``Quantum Chaos and Holographic Tensor Models,''
  JHEP {\bf 1703}, 056 (2017)
  doi:10.1007/JHEP03(2017)056
  [arXiv:1612.06330 [hep-th]].
  

 
 

 
 
}
\end{thebibliography}
\end{document}